%% file: ANA-PIXE-2019-02-PAPER.tex
\documentclass[a4paper,11pt]{article}
\pdfoutput=1 

\usepackage{jinstpub} 
\usepackage{siunitx}
\usepackage{overpic}

\newcommand*{\e}{$e^-$}                
\newcommand*{\lumi}{\centi\meter$^{-2}$\second$^{-1}$}
\DeclareSIUnit \nq {\text{n}_{\text{eq}} \per \centi \meter ^{2}}
\DeclareSIUnit \lumi {\centi\meter^{-2}\second^{-1}}
\DeclareSIUnit \xzero {\%\,X_0}
\DeclareSIUnit \e {e^-}

\newcommand*{\barn}{b}

\title{\boldmath Measurements of Single Event Upset in ATLAS IBL}

\author[2a]{G.~Balbi,}
\author[3,11]{M.~Barbero,}
\author[8a,101]{R.~Beccherle,}
\author[9]{M.~Bindi,}
\author[11]{P.~Breugnon,}
\author[4,14]{P.~Butti,}
\author[6]{D.~Cinca,}
\author[1]{J.~Dickinson,}
\author[7]{D.~Ferrere,}
\author[11]{D.~Fougeron,}
\author[1]{M.~Garcia-Sciveres,}
\author[16]{J.~Garcia~Pascual,}
\author[8a,8b]{A.~Gaudiello,}
\author[8a]{C.~Gemme,}
\author[2a,2b]{N.~Giangiacomi,}
\author[3]{T.~Hemperek,}
\author[1, 17]{L.~Jeanty,}
\author[13]{O.~Kepka,}
\author[14]{M.~Kocian,}
\author[3]{K.~Lantzsch,}
\author[1,16]{P.~Liu,}
\author[12]{C.~Martin,}
\author[1]{A.~Mekkaoui,}
\author[11]{M.~Menouni,}
\author[1,5]{K.~Potamianos,}
\author[11,*]{A.~Rozanov,}
\author[10]{Y.~Takubo,}
\author[15]{and M.~Wensing.}

\affiliation[1]{Physics Division, Lawrence Berkeley National Laboratory and University of California, Berkeley CA, United States of America}

\affiliation[2a]{INFN Sezione di Bologna, Bologna, Italy}

\affiliation[2b]{Dipartimento di Fisica, Universit\`a di Bologna, Bologna, Italy}

\affiliation[3]{Physikalisches Institut, University of Bonn, Bonn, Germany}

\affiliation[4]{CERN, Geneva, Switzerland}

\affiliation[5]{DESY, Hamburg and Zeuthen, Germany}

\affiliation[6]{Fakult\"{a}t Physik, Technische Universit\"{a}t Dortmund, Dortmund, Germany}

\affiliation[7]{Section de Physique, Universit\'e de Gen\`eve, Geneva, Switzerland}

\affiliation[8a]{INFN Sezione di Genova, Genova, Italy}

\affiliation[8b]{Dipartimento di Fisica, Universit\`a di Genova, Genova, Italy}

\affiliation[9]{II Physikalisches Institut, Georg-August-Universit\"{a}t, G\"{o}ttingen, Germany}

\affiliation[10]{KEK, High Energy Accelerator Research Organization, Tsukuba, Japan}

\affiliation[11]{CPPM, Aix-Marseille Universit\'e and CNRS/IN2P3, Marseille, France}

\affiliation[12]{Ohio State University, Columbus OH, United States of America}

\affiliation[13]{Institute of Physics, Academy of Sciences of the Czech Republic, Praha, Czech Republic}

\affiliation[14]{SLAC National Accelerator Laboratory, Stanford CA, United States of America}

\affiliation[15]{Fakult{\"a}t f{\"u}r Mathematik und Naturwissenschaften, Bergische Universit\"{a}t Wuppertal, Wuppertal, Germany}

\affiliation[16]{Institute of High Energy Physics, Chinese Academy of Sciences, Beijing, China}

\affiliation[17]{Institute for Fundamental Science, University of Oregon, Eugene, OR, USA}

\affiliation[101]{now at INFN Sezione di Pisa, Pisa, Italy}

\affiliation[*]{Corresponding author}

\emailAdd{rozanov@cppm.in2p3.fr}

\abstract{Effects of Single Event Upsets (SEU) and Single Event Transients (SET) are studied
 in the FE-I4B chip of the innermost layer of the ATLAS pixel system. 
SEU/SET affect the FE-I4B Global Registers as well as the settings for
the individual pixels, causing, among other things, occupancy losses,
drops in the low voltage currents,
noisy pixels, and silent pixels.
 Quantitative data analysis and simulations indicate that SET dominate over SEU on the load line of the memory.
 Operational issues and mitigation techniques are presented.}


\begin{document}
\maketitle
\flushbottom


\section{Introduction}
\label{sec:intro}
\input{documents/introduction}

\section{The ATLAS IBL detector}
\label{sec:ibldetector}
\input{documents/ibldetector}

\section{Global and pixel configuration memories}
\label{sec:config}
\input{documents/configuration}

\section{SEU measurements}
\label{sec:seumeas}
\input{documents/seumeasurements}

\section{Read-back measurements}
\label{sec:readback}
\input{documents/readback}


\section{Recovery of bits flipped by SEU}
\label{sec:reconfig}
\input{documents/recovery}


\section{Conclusion}
\label{sec:conclusion}
\input{documents/conclusion}




\appendix
\section{Description of IBL detector notations}
In this appendix we will give a short description of the IBL detector in terms of geometric layout, power system (LV and HV) and data acquisition (DAQ) granularities. A full description can be found elsewhere \cite{ref:IBLPaper}.

To indicate a specific IBL DAQ module, the following name scheme is adopted:

\begin{center}
LI\_S\textbf{XX}\_A/C\_M\textbf{YY}\_A/C\textbf{ZZ}
\end{center}
where \begin{itemize}
\item \textbf{XX} refers to the stave ID number,
\item A/C refers to the position of the module in the longitudinal axis with A $\rightarrow z > 0$ and C $\rightarrow z < 0$,
\item \textbf{YY} refers to the DCS module group and goes from 1 to 4 depending on the position of the module along the stave,
\item \textbf{ZZ} refers to the DAQ module that goes from 1 to 8 depending on the position of the module along the stave.
\end{itemize}

\begin{figure}
  \centering
   \begin{overpic}[width=1\linewidth]{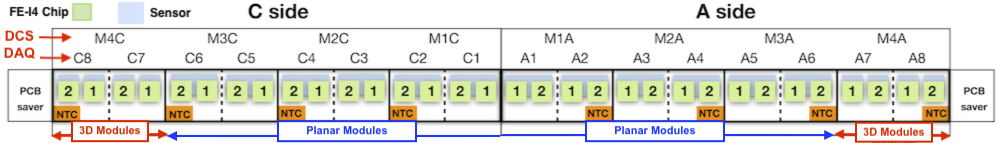}
  \end{overpic}
  \caption{Schematic for a full stave of the module and FE-I4B layout, as well as the DCS/power system modularity. The FE-I4B chips are labelled as A(C)X-Y where X runs from 1 to 8 towards the high $|\eta|$ direction, and Y is 1 or 2. $A(C)7-1/2$ and $A(C)8-1/2$ are single-chip 3D modules. The DCS modularity is shown as M1A(C) through M4A(C). The NTCs measure the temperature of each DCS module group.}
\label{fig:IBLStaveModuleName}
\end{figure}

A graphical representation of the naming scheme is shown in Fig. \ref{fig:IBLStaveModuleName}.

One example of naming scheme for a DAQ module is LI\_S\textbf{02}\_C\_M\textbf{4}\_C\textbf{8}

In case a single FE needs to be addressed, an extra \textbf{1/2} is append to the end of the  module name.

One example of the naming scheme for a single FE is LI\_S\textbf{02}\_C\_M\textbf{4}\_C\textbf{8}-\textbf{2}.

In this document, DCS module groups (powering 4 FEs with the same LV and HV), DAQ modules (controlling and reading out 2 FEs), Sensors (connected to 2 or 1 FEs depending on the planar or 3D technology) or single FEs will be used depending on the granularity affected.

\acknowledgments



\input{acknowledgements/ANA-PIXE-2019-02-PAPERAcknowl}




\end{document}

%% file: documents/introduction.tex
During operation, electronics at the Large Hadron Collider (LHC) experiments must contend with radiation damage both from integrated dose and from instantaneous Single Event Effects (SEE)~\cite{ref:softErrors}. The most prevalent SEE is the flipping of a logic state, typically in a memory cell, known as Single Event Upset (SEU). The readout integrated circuits in the ATLAS pixel detector were custom designed with SEU-hardened memory cells~\cite{ref:TWEPP2012}
(Dual Interlocked CElls or DICE latches~\cite{ref:DICE} and triple redundancy). These reduce the SEU rate, but do not completely eliminate it. 
Another frequent SEE is the creation of short pulses, called Single Event Transient (SET) or glitches. These glitches may propagate through the chip
and flip the logic state of the memory.

A sufficiently large amount of charge must be injected into the sensitive nodes of a circuit to flip its logic states. This corresponds to an ionization threshold. Even for memory cells that are not hardened, this threshold is much higher than the specific ionization of minimum ionizing particles (MIPs). Thus, at the LHC, SEUs are not caused by the MIPs that the detectors were built to measure, but by highly ionizing recoil nuclei and by showers resulting from occasional nuclear interactions of the MIPs in the proximity of the memory cells~\cite{ref:ropp2018}.
These secondary interactions are complex to model, and it is therefore interesting to measure SEU rates in-situ and compare 
these rates to test beam results acquired during the detector construction phase.
  It is also 
important 
to quantify the operational consequences of SEUs in FE-I4B.

In this paper we analyze data from operation of the ATLAS innermost pixel layer, called the Insertable B-Layer (IBL)~\cite{ref:IBL}-\cite{ref:IBLPaper}.
We  quantify the influence of SEU on the pixel matrix and on the data readout, and extract SEU cross sections that can be compared to test beam data.  The results have led to improvements in operation of the present detector and mitigation of SEU effects, and they provide an important reference for the design of future detectors.

\begin{figure}
\centering
\begin{overpic}[width=0.6\linewidth]{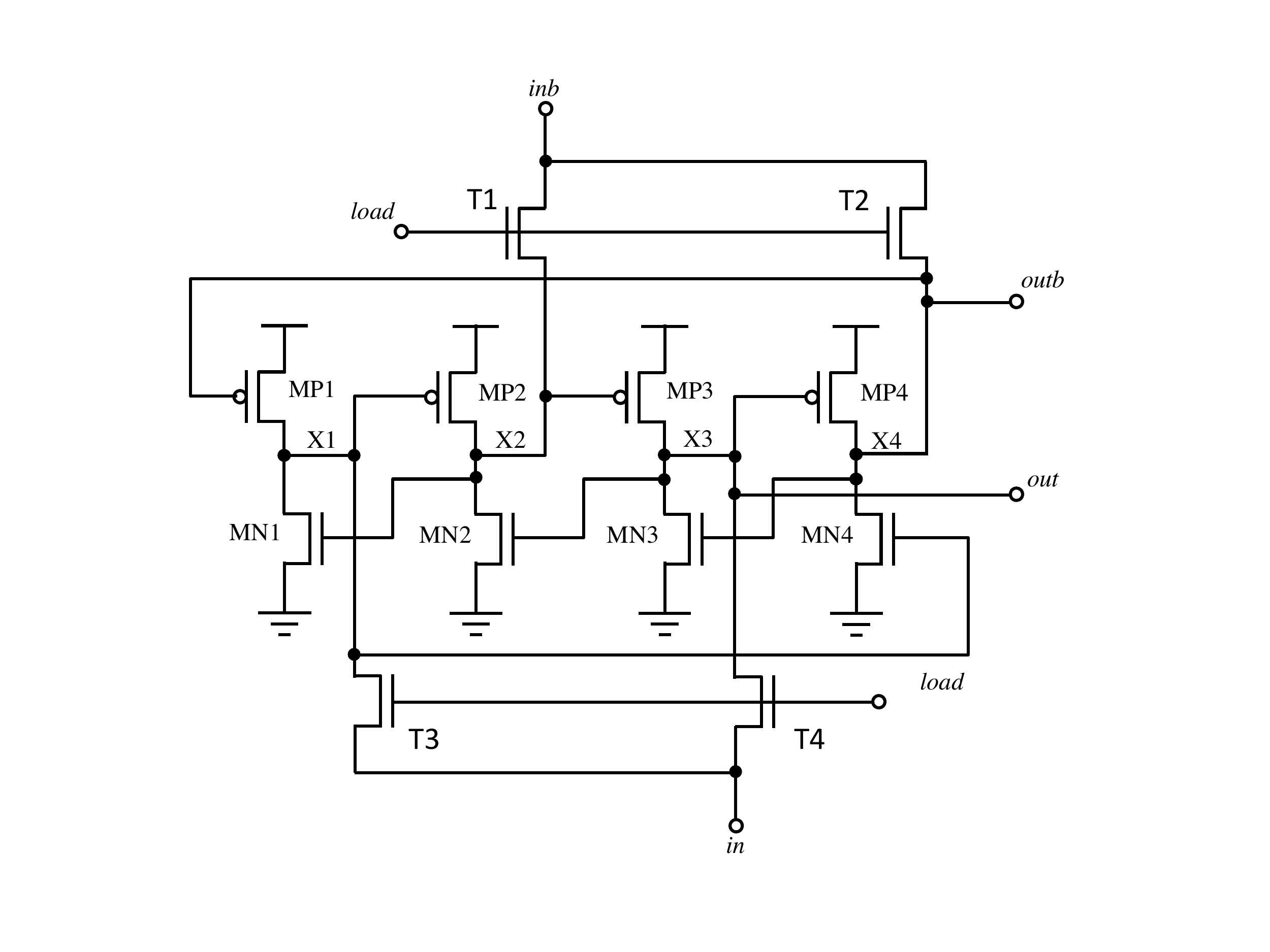}
\end{overpic}
\caption{DICE latch structure, based on the conventional cross coupled inverter latch
 structure where the four nodes X1 to X4 store data as two pairs of complementary
 values. The original cell state is restored when an SEU error is introduced
in a single node.}
\label{fig:dice_scematics}
\end{figure}

\begin{figure}
\centering
\begin{overpic}[width=0.6\linewidth]{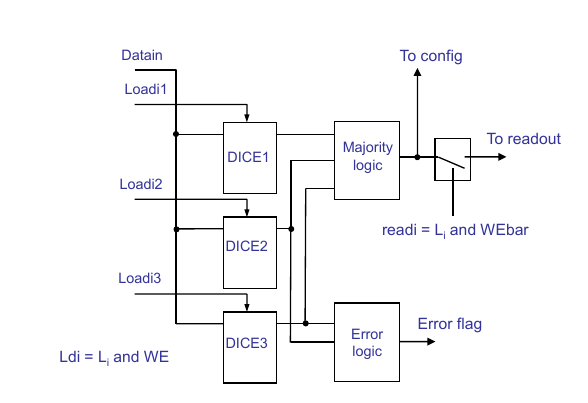}
\end{overpic}
\caption{The FE-I4B global memory cell is based on the triplication of DICE latches with the
addition of the simple majority logic. No automatic error correction is used,
but single bit flips can be monitored by an error counter. The memory has two
 independent ports. The first (Datain, Load) is used to write data coming from
 the command decoder and the second (To-config, To-readout) to read the
 majority logic output. The output data bit is corrupted only when more than
 one DICE latch is upset.}
\label{fig:global_memory}
\end{figure}

Single DICE latch memory cells (see figure~\ref{fig:dice_scematics}) are used to store pixel configuration values, while triple redundant DICE latches (see figure~\ref{fig:global_memory}) are used to store chip-wide (or global) configuration values.

The layout at transistor level is optimized to maximize distance between coupled nodes and between redundant latches, for maximal reduction of upset rates.
Upset of the specific pixel configuration latch can usually only affect a single pixel at a time, while upset of global configuration memory can impact many pixels and even the operation of the entire chip.
It is possible to directly detect upsets in the global configuration, but not upsets in the local pixel configuration.
While it is possible to read back pixel latch values (see section~\ref{sec:readback}),
this requires special operations not typically performed during data taking.
Different techniques have been applied to measure pixel latch updates in physics data without directly reading out the stored values.
These involve understanding the effect of flipping each configuration bit and searching for these effects in data. The main effects
studied are called quiet pixels (when a pixel's efficiency drops abruptly, see section~\ref{sec:quiet})
and noisy pixels (when a pixel suddenly fires at a high rate, see section~\ref{sec:noisy}).
In special runs, pixels can be deliberately configured to begin in the off state,
 and a search is performed with the aim of monitoring  when they become active, see section~\ref{sec::LHCfill}.

\begin{figure}[h!]
\centering
\begin{overpic}[width=0.6\linewidth]{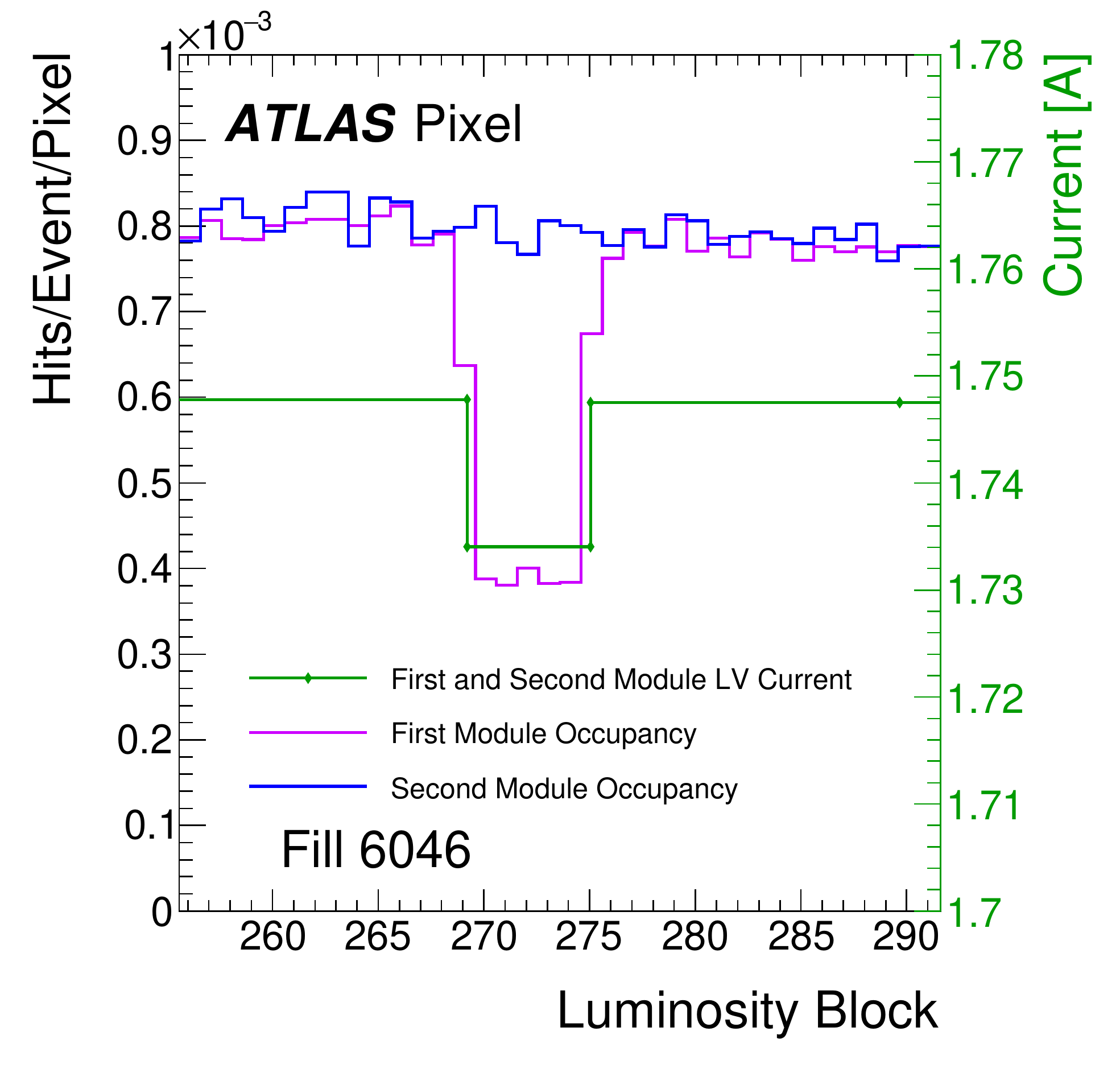}
\end{overpic}
\caption[Caption with footnote]{Effects of SEU on FE-I4B global registers can be seen during a typical LHC fill; in this case (fill 6046), the peak luminosity reached is $1.5 \times 10^{34}$\,cm$^{-2}$\,s$^{-1}$, and  \SI{490}{\per\pico\barn} were delivered over the entire fill. During the data taking, at luminosity block \footnotemark[1] (LB) $\sim\,268$, a drop in the LV current consumption can be observed (green line). At the same time, a drop in occupancy is observed in one (pink line) of the two DAQ modules that share the same LV power supply (see the Appendix for details about the DAQ and power system granularity).  At LB $\sim\,277$, the DAQ module was manually reconfigured, bringing the LV current consumption and hit occupancy back to their values before the SEU.} 
\label{fig:SEUEffectsGlobal}
\end{figure}

During 2017, the LHC peak instantaneous luminosity increased with respect to 2016 and was constantly above $1.5 \times 10^{34}$\,cm$^{-2}$\,s$^{-1}$. Under these conditions, the effect of SEUs on the FE-I4B global registers was greater than in previous years, and more frequent front end chip reconfigurations were needed to preserve good data quality and data taking efficiency. The effects of SEUs were visible in the behavior of the FEs: impacted FEs can stop sending hits, become very noisy, or experience large drops/increases
(up to \SI{\pm 100}{\milli\ampere}) of the low voltage (LV) current consumption monitored from the Detector Control System (DCS) (see figure~\ref{fig:SEUEffectsGlobal}). 
Starting in 2016, some SEU/SET effects were addressed by manual or automatic (``Quick Status'') reconfiguration
of the affected modules.

To minimize the impact of SEUs on ATLAS data taking, a procedure was implemented to regularly reconfigure the global registers of the FE-I4B readout chips in the entire IBL.
Thanks to a joint effort of online software and firmware (see section~\ref{subsec:GRRecon}), it was possible to introduce this procedure without any additional dead time in ATLAS. Starting in August 2017, the global registers of the IBL FE-I4B chips were reconfigured approximately every five seconds, improving the overall Data Acquisition (DAQ) efficiency and eliminating the low voltage current fluctuations that were previously observed. Unfortunately it was not possible to also regularly reconfigure the single pixel DICE latches in the FE-I4B: the software modifications needed impacted the overall stability of the DAQ system. However, a test run was performed in July 2018 and can be used as proof of concept for future implementations (see details in section~\ref{subsec:PRRecon}).

\interfootnotelinepenalty=10000
\footnotetext[1]{Luminosity Block (LB) is a time interval of ATLAS data recording over which the experimental conditions are assumed to be constant. LB duration is usually around one minute.}

%% file: documents/ibldetector.tex
\begin{figure}
  \centering
  \begin{overpic}[width=0.7\linewidth]{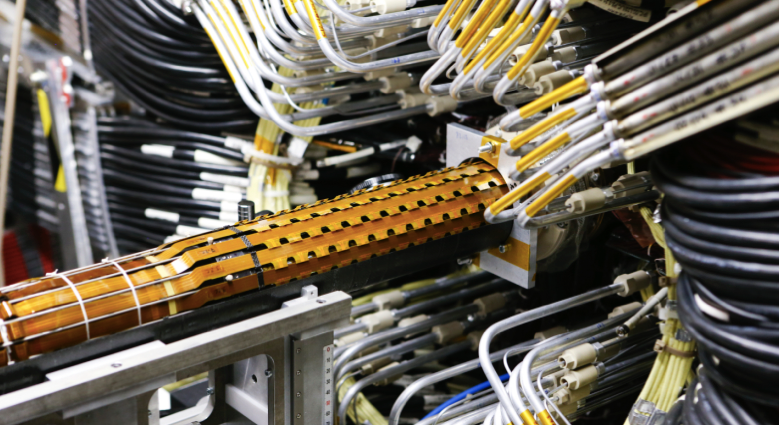}
  \end{overpic}
  \caption{Image of the IBL being inserted (from left to right) into the pre-existing ATLAS Run 1 Pixel detector. A typical barrel structure divided in staves along the $\phi$ coordinate is clearly visible in the picture. }
\label{fig:IBL}
\end{figure}

The original ATLAS Run 1 Pixel detector was upgraded for Run 2 with the addition of the IBL, which provides an additional layer of pixels at $\sim 3.4$ cm from the interaction point.
The IBL greatly improves impact parameter resolution, vertexing, b-tagging and tracking robustness after high radiation doses.
The IBL detector consists of one layer of pixel sensors (see figure~\ref{fig:IBL}) built around the beam pipe, providing full azimuthal ($\phi$) hermeticity for high transverse momentum ($p_T > 1$ GeV) particles and longitudinal coverage up to $|\eta| = 3 $. Two different sensor types were chosen: a planar design in the central area ($|\eta| < 2.5$) based on the already existing ATLAS Pixel sensor, and 3D sensor designs in the forward regions ($2.5 < |\eta| < 3$) which are applied here for the first time in a high energy physics experiment.

\begin{figure}
\centering
\begin{overpic}[width=0.5\linewidth]{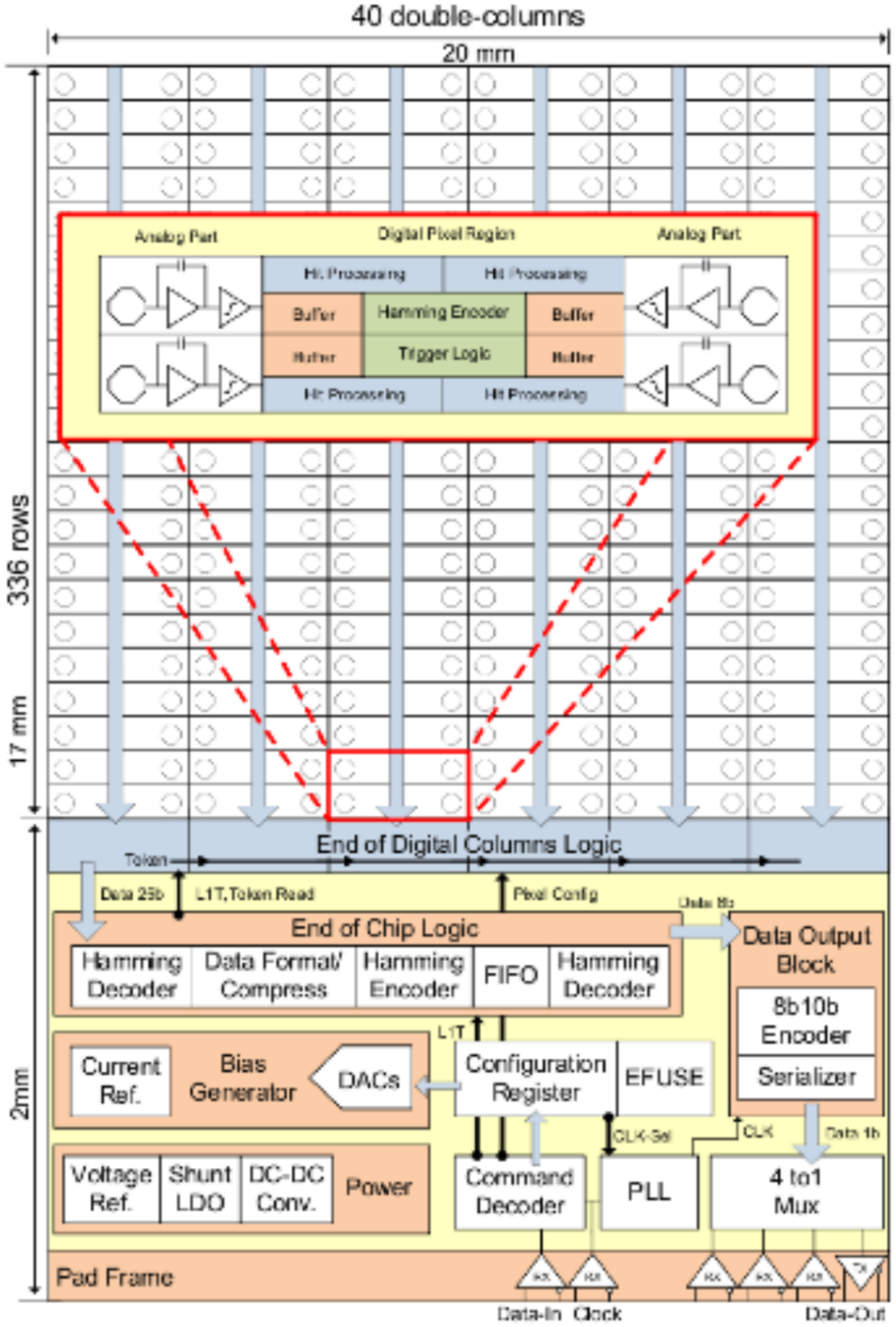}
\end{overpic}
\caption{Diagram of the FE-I4B chip.}
\label{fig:fei4bdiagram}
\end{figure}

The IBL uses the FE-I4B readout chip~\cite{ref:FE-I4},~\cite{ref:FE-I4guide}, which is built in \SI{130}{\nano\meter} CMOS technology.
 A top level diagram of the FE-I4B is shown in figure~\ref{fig:fei4bdiagram}.
Pixel sensor with each pixel nominal size of $250 \times 50 ~\mu$m$^2$ is bonded to FE-I4B chip. On the edges of the chip there are long pixels with the size of
$500 \times 50 ~\mu$m$^2$ or $450 \times 50 ~\mu$m$^2$ ~\cite{ref:IBLPaper}.

\begin{figure}
  \centering
  \begin{overpic}[width=0.5\linewidth]{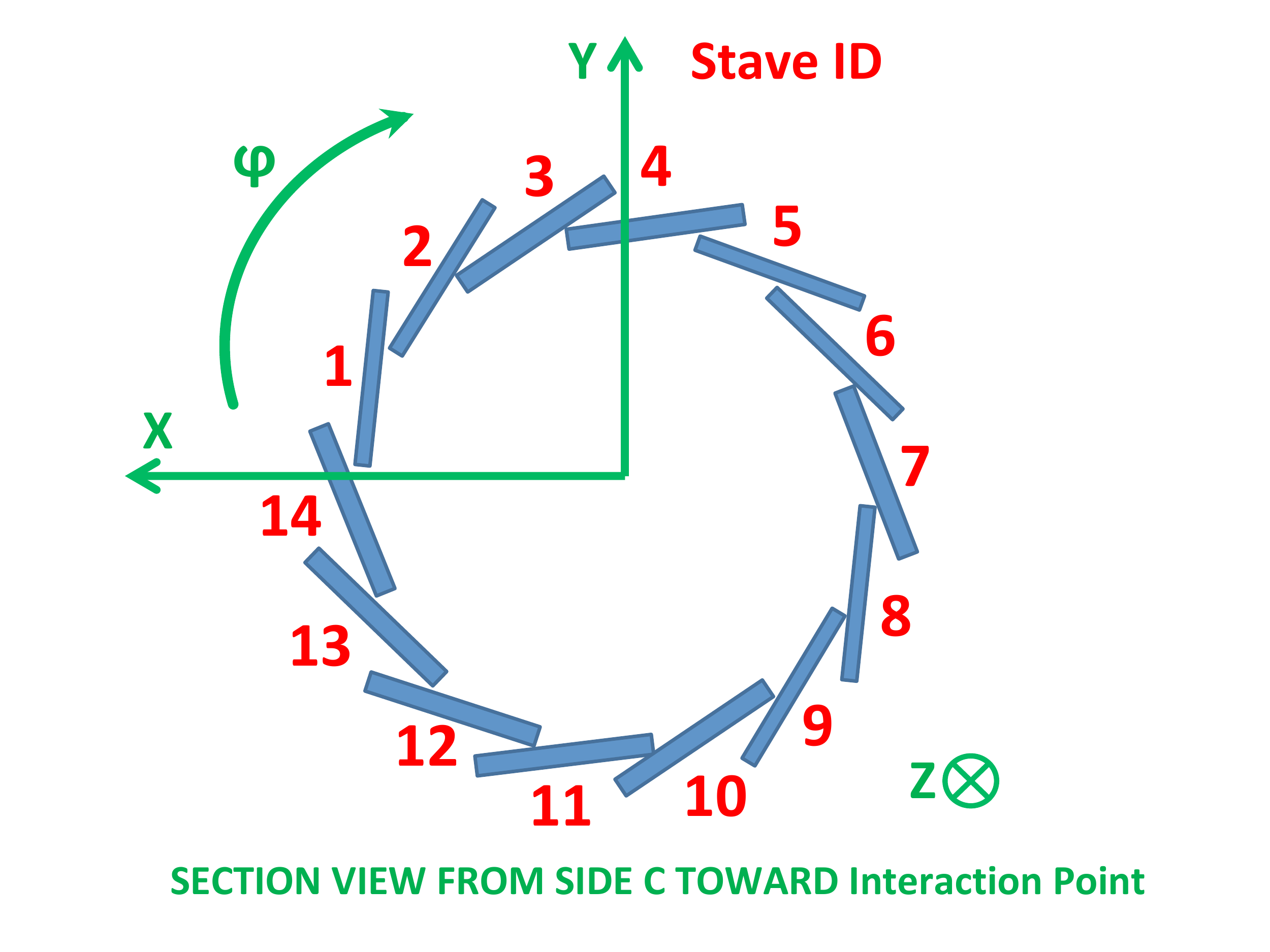}
  \end{overpic}
  \caption[footNote]{Schematic of the IBL detector, transverse view ($x-y$ plane). The concentric IBL staves (typically numbered from 1 to 14 following the stave position index) are oriented around the beam pipe along the $\phi$ coordinate.\footnotemark[2]}
\label{fig:IBLWheel}
\end{figure}

The detector is made of 14 carbon fiber staves, 64.3 cm long and 2 cm wide. A transverse view of the IBL staves is shown in figure~\ref{fig:IBLWheel}.
The Stave ID numbering scheme counts from 1 to 14, increasing with the $\phi$ angle.

\begin{figure}
  \centering
  \begin{overpic}[width=0.9\linewidth]{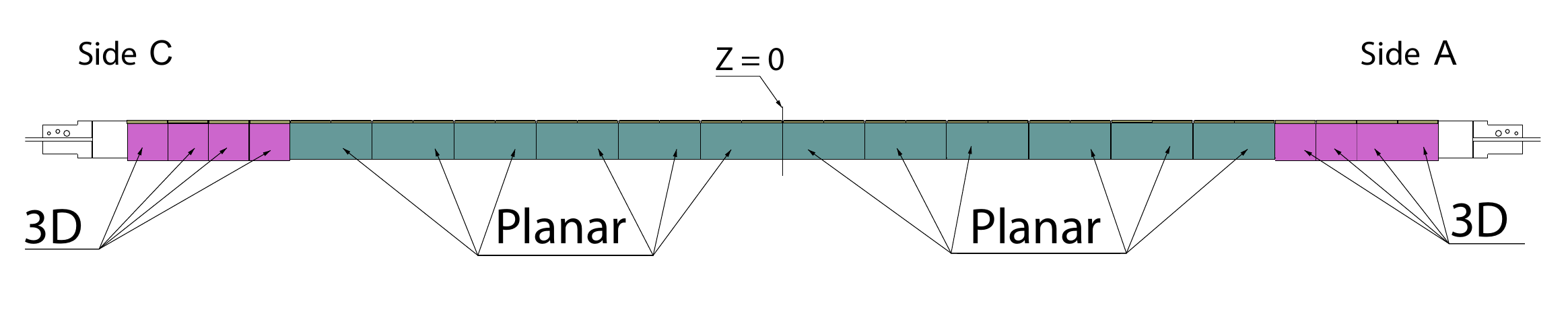}
  \end{overpic}
  \caption{Schematic of an IBL stave, longitudinal view (along the $z$ axis). The central modules (read out via double-FE-I4B chips) use planar sensor technology, while the forward modules (read out via single-FE-I4B chips) use 3D sensor technology.}
\label{fig:IBLStaveStructure}
\end{figure}

Each stave hosts 20 pixel sensor modules (see figure~\ref{fig:IBLStaveStructure}) corresponding to 12 double-FE-I4B chip modules (for the planar sensors) and 8 single-FE-I4B chip modules (for the 3D sensors).
In total, there are 32 FE-I4B chips per stave, equivalent to 448 FE-I4Bs and ca. 12 million pixels to be read out for the entire IBL.

\footnotetext[2]{The nominal interaction point is defined as the origin of the coordinate system, while the beam direction defines the $z$-axis such that
$z= x \times y$ with $x$ pointing to the center of LHC and $y$ upwards. 
The azimuthal angle $\phi$ is measured around the beam axis, and the polar angle $\theta$ is the angle from the beam axis.}

\begin{figure}
\centering
  \begin{overpic}[width=0.8\linewidth]{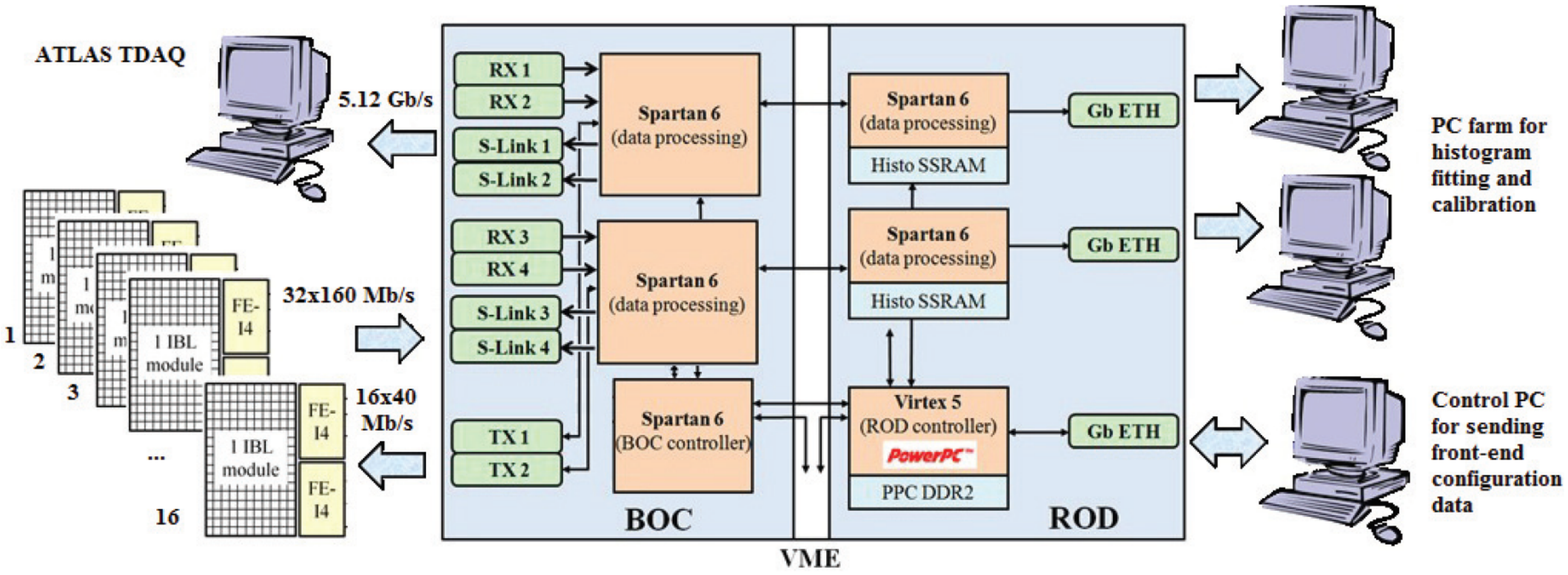}
\end{overpic}
\caption{Schematic view of the IBL readout system. From left to right: the detector, the BOC, the ROD and the calibration farm. All
detector readout fibers are located at the back of the crate (BOC). New user-pluggable connections (GBit Ethernet) are on the front
side (ROD).}
\label{fig:IBLReadout}
\end{figure}

The IBL readout electronics are designed to exploit 
 the enhanced detector performance and overcome the limitations present in ATLAS Pixel Run 1 system. The IBL readout system consists of two 9U-VME cards: the Back-of-Crate (BOC) and Read-Out Driver (ROD) which implement optical I/O interface and data processing, respectively (see figure~\ref{fig:IBLReadout}). Each card pair processes data received from a full IBL stave, corresponding to 32 FE-I4B data links working at 160 Mb/s, for a total I/O bandwidth of 5.12 Gb/s. Moreover, the new ROD design implements dedicated Gigabit Ethernet connections, thus enhancing the overall acquisition performance of the calibration runs. 
Configuration and control of the read-out electronics are the main tasks of the ROD card. In order to have new capabilities, a PowerPC (PPC) was embedded into the main ROD FPGA, a Xilinx Virtex-5 (see figure~\ref{fig:IBLReadout}). The PPC solution enhanced the control and monitoring performance; in particular, a specific software task to recover the FE-I4B from SEU was developed and deployed in the system during Run 2.

Further details about the IBL detector layout and naming convention
are covered in the Appendix, including the naming schemes used for the
power system, the data-acquisition modules, the sensors, and the front-end chips.

%% file: documents/configuration.tex
A 13 bit register is available to configure each pixel: it is comprised of a 1-bit enable flag, a 5-bit threshold tuning DAC (TDAC), a 4-bit time-over-threshold (ToT) tuning DAC (FDAC), a 1-bit HitBus (input to logical OR of all pixel discriminator outputs in the matrix), and 2 bits for the selection of the charge injection capacitor.
The ToT represents the length of time that single pixel discriminator is over threshold. It has a 4-bit resolution, in counts of an externally supplied clock, nominally 40 MHz, that corresponds to the LHC bunch crossing (BC) time of 25 ns.

\begin{figure}[h]
\centering
 \begin{overpic}[width=0.95\linewidth]{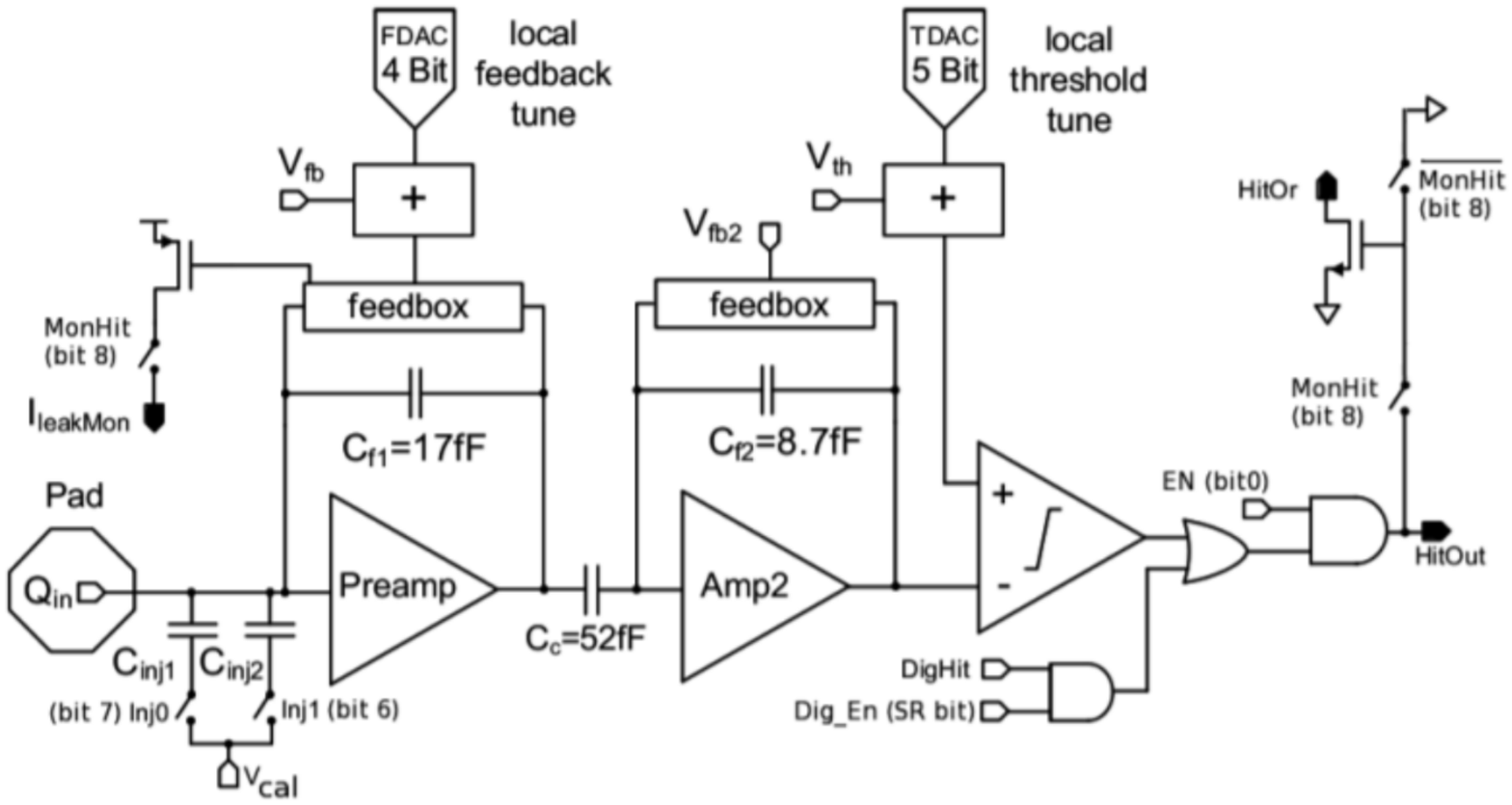}
\end{overpic}
\caption{Diagram of the FE-I4B analog pixel.}
\label{fig:analogpixeldiagram}
\end{figure}

The pixel analog front end of FE-I4B is shown schematically in figure~\ref{fig:analogpixeldiagram}. Single DICE latches are used to store bits within each pixel.
 
In data taking configuration, the output enable bit mostly stores ones, as most of the pixels are enabled. However, a small fraction of pixels (tenths of a percent) corresponding to noisy pixels are disabled during calibration. 
The TDAC value is centered around fifteen, and the FDAC value is typically centered around seven. 
The capacitor selection bits are both set to one, and the HitBus flag is set to one (meaning HitOr disabled).

The Global Configuration Memory, located at the end of the column area outside of the pixel matrix region, is implemented as a memory block of 32 words of 16 bits (512 bits in total). The design used for this global memory is based on the triplication of the DICE latches to further suppress SEU. Such triplication is not possible inside the pixel due to space constraints. 
The global threshold, generated by a coarse and a fine DAC, is vital for proper chip functionality. 
Together the two DACs give a unique Global DAC (GDAC) value that is typically tuned, along with local TDAC registers, to achieve a target value of \SI{2500}{\e} threshold for each pixel.

The occurrence of SEUs during data taking modifies both single pixel and global configurations, producing quiet pixels, noisy pixels, and introducing a general detuning of the FE-I4.


%% file: documents/seumeasurements.tex
\label{sec:seumeasurements}
The SEU cross section is equal to the total number of bit flips observed in the whole memory 
($N_{\textrm{flips}}$)
over the proton fluence achieved during the test ($\Phi$), and divided by the number of latches  implemented in the chip
($N_{\textrm{latches}}$):
\vspace{-0.5cm}
\begin{center}
\begin{equation}
\label{eq:cs}
\sigma=\frac{N_{\textrm{flips}}}{\Phi\cdot N_{\textrm{latches}}}.
\end{equation}
\end{center}

\subsection{Beam tests}
\label{subsec:beamtests}
Irradiation tests were carried out in 2012 using the IRRAD3 beam line of the Proton Synchrotron (PS) Irradiation Facility~\cite{ref:IRRAD} at CERN. The IRRAD3 provides a beam of 24\,GeV protons, and the structure of the beam is defined by the operation cycle of the PS accelerator. 

During the irradiation testing, devices were oriented perpendicular
 to the radiation beam line. 
The enable bits of all pixels were disabled before the test beam. The $0\rightarrow1$ transition of the pixel enable bit causes a pixel to respond to charge injection. The test beam was periodically injected and the number of active pixels was measured.
The SEU cross section measured in the 2012 beam test was $1.1\times 10^{-15}$\,cm$^{2}$~\cite{ref:TWEPP2012} for a DICE latch
implementation (FE-I4A, columns 60-75) that closely matches the implementation 
in FE-I4B.

\subsection{Analysis of cluster data from physics running}
\label{subsec:clusterdata}
SEU effects can flip the enable bit of a pixel from 1 to 0, making the pixel quiet. The SEU flip of a TDAC bit can lower the pixel threshold and make the pixel noisy. Therefore, the sensitivity of DICE latches to SEU can be estimated by studying such quiet or noisy pixels.

\begin{figure}[h]
\centering
\begin{overpic}[width=0.82\linewidth]{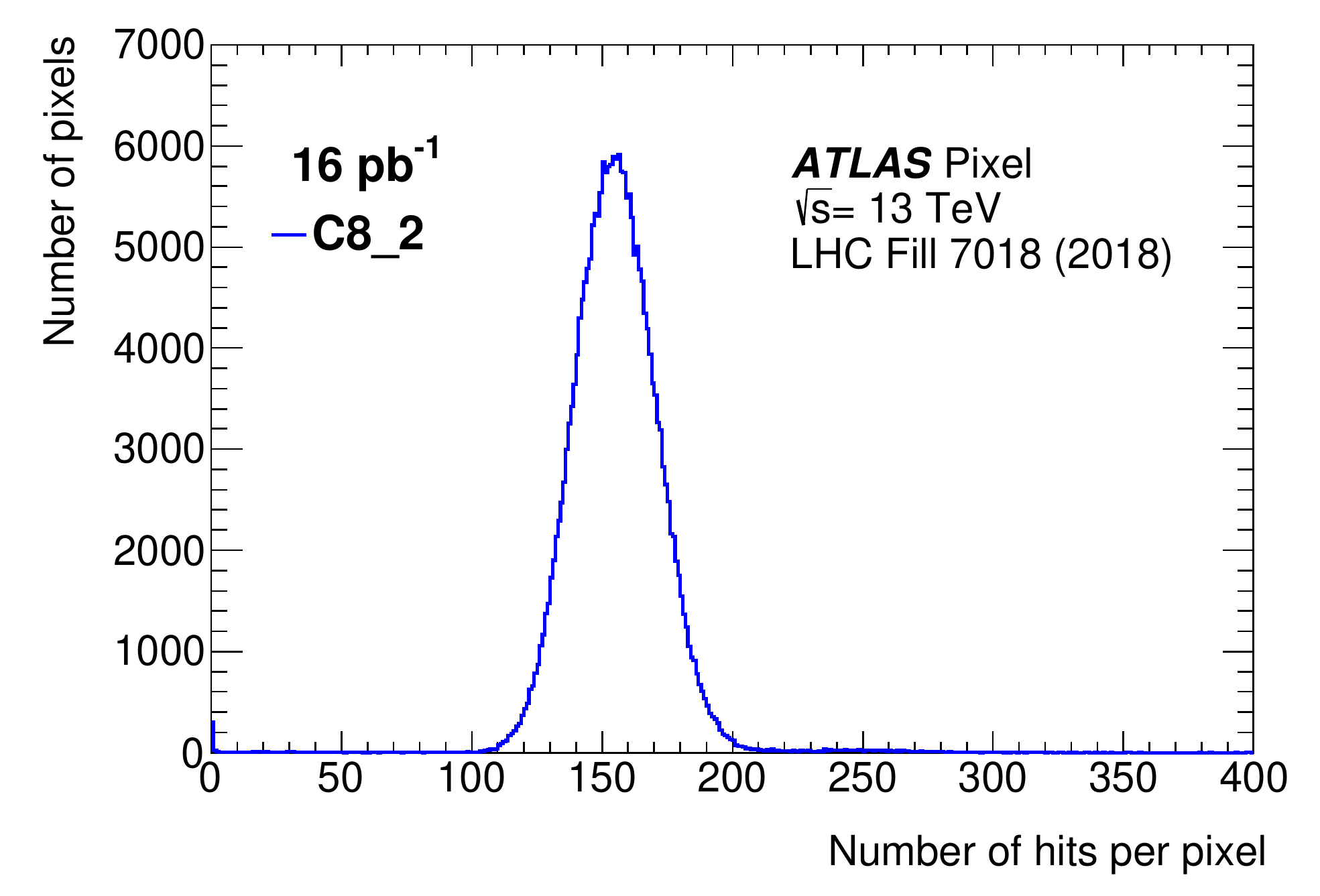}
\end{overpic}
\vspace{0.15cm}
\caption{Occupancy of each pixel in the first \SI{16}{\per\pico\barn} of 
data of LHC fill 7018 from 2018, shown in the most forward IBL ring on the negative-$z$ side. 
}
\label{fig:fill5163-2016a}
\end{figure}

\begin{figure}[h]
\centering
  \begin{overpic}[width=0.82\linewidth]{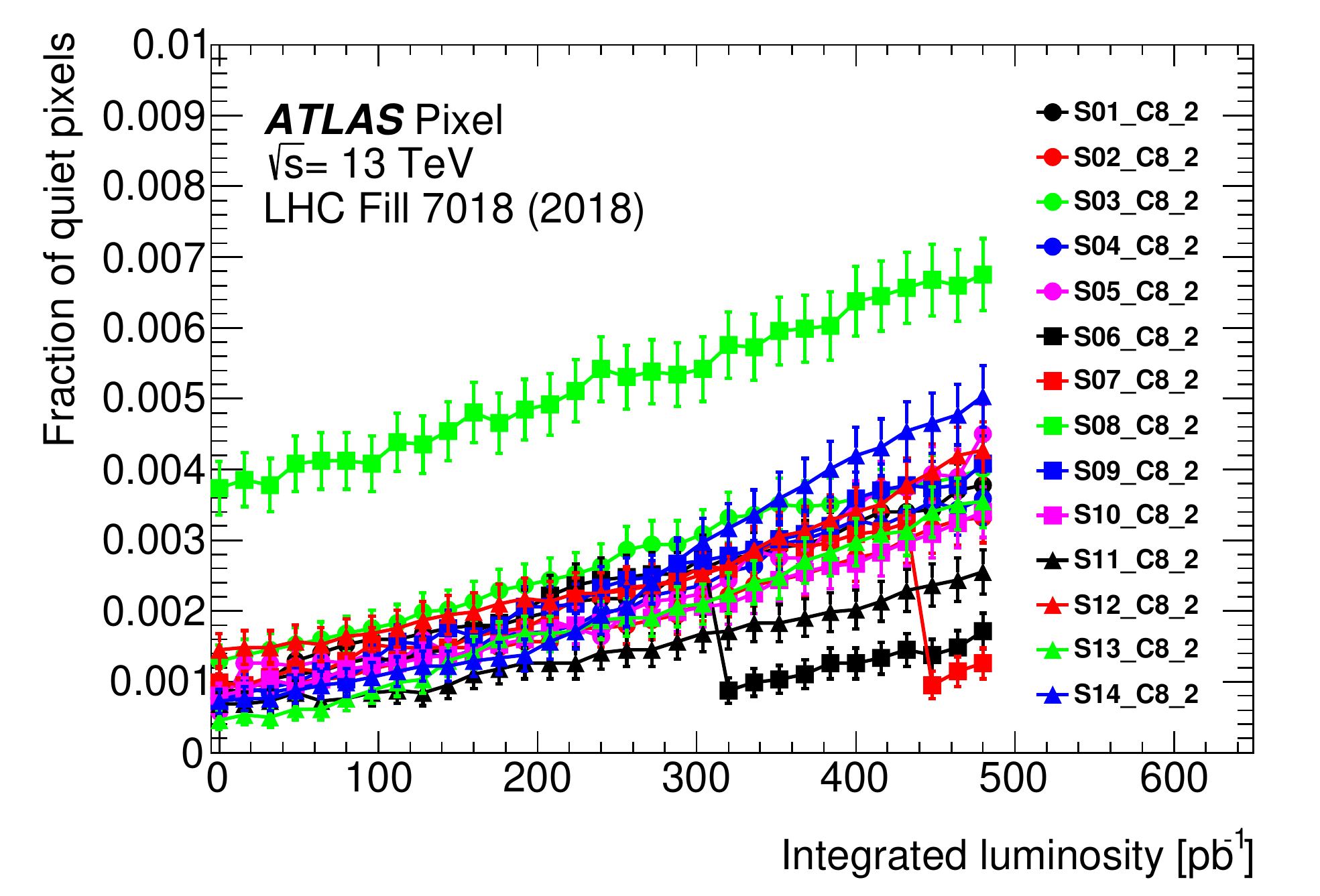}
\end{overpic}
\vspace{0.15cm}
\caption{Fraction of quiet pixels versus integrated luminosity in the fill 7018, shown in all 14 modules in the most forward IBL ring on the negative-$z$ side. The 14 modules are displaced clockwise around the beam pipe with the first one at $\phi$=\ang{-6.1}. Long pixels were removed from the analysis.}
\label{fig:fill5163-2016b}
\end{figure}

Figure~\ref{fig:fill5163-2016a} shows the occupancy of each pixel in one 3D module in the most forward IBL ring (with a width of one FE-I4B chip) 
on the negative-$z$
 side. The data are extracted from randomly triggered events in the first \SI{16}{\per\pico\barn} of LHC fill 7018 in 2018. 
Long pixels, which have more hits than normal pixels, are not included in the sample. The average occupancy of one pixel in \SI{16}{\per\pico\barn} of data from this fill is about 155.

\subsubsection{Quiet pixels \label{sec:quiet}}
 A pixel is defined to be quiet if it fired zero times in \SI{16}{\per\pico\barn} of data taking.
Figure~\ref{fig:fill5163-2016b} shows the fraction of quiet pixels versus integrated luminosity ($\mathcal{L}$) in the 14 modules in the most forward IBL ring on the negative-$z$ side. The number of quiet pixels increases during data taking due to the accumulation of pixels with a flipped enable bit.
In two modules one can observe a fast drop in the number of quiet pixels
due to the reconfiguration of these modules.
Module-to-module differences in the initial number of quiet pixels indicate different fractions of silent pixels, which fire zero times during the entire fill. The fraction of pixels that become quiet due to SEU is seen to increase linearly with integrated luminosity.
 The dependence of the number of quiet pixels on luminosity
was fitted 
 with a linear function $p_0+p_1\cdot\mathcal{L}$,
where the mean $p_1$ is  $(5.4 \pm 1.3) \times10^{-3}$ fb.

The fraction of pixels that become quiet due to SEU ($p_1\cdot\mathcal{L}$) is equal to the ratio $\frac{N_{\textrm{errors}}}{N_{\textrm{latches}}}$ in Eq.~\ref{eq:cs}.  The pixel latch SEU cross section in FE-I4B is calculated with the ``quiet-pixels-fraction":
\vspace{-0.5cm}
\begin{center}
\begin{equation}
\label{eq:cs-new}
\sigma=\frac{p_1\cdot\mathcal{L}}{\Phi}
\end{equation}
\end{center}

The predicted flux of hadrons with energy above 20 MeV
with PYTHIA/FLUKA ~\cite{ref:PYTHIA}-\cite{ref:FLUKA} simulations in the extreme outside 3D sensor IBL module
is ${\Phi}= 0.91\times10^{13}$ hadrons
(energy above 20 MeV) cm$^{-2}$ per \SI{1}{\per\femto\barn}  ~\cite{ref:IanDawson}.
 The SEU cross section is calculated to be 
$(0.60 \pm 0.14) \times10^{-15}$cm$^{2}$,
which is of the same order of magnitude as the test beam result.

\subsubsection{Z-dependence of appearance of quiet pixels}~\label{sec::zdependence}
\begin{figure}[h]
\centering
 \begin{overpic}[width=0.82\linewidth]{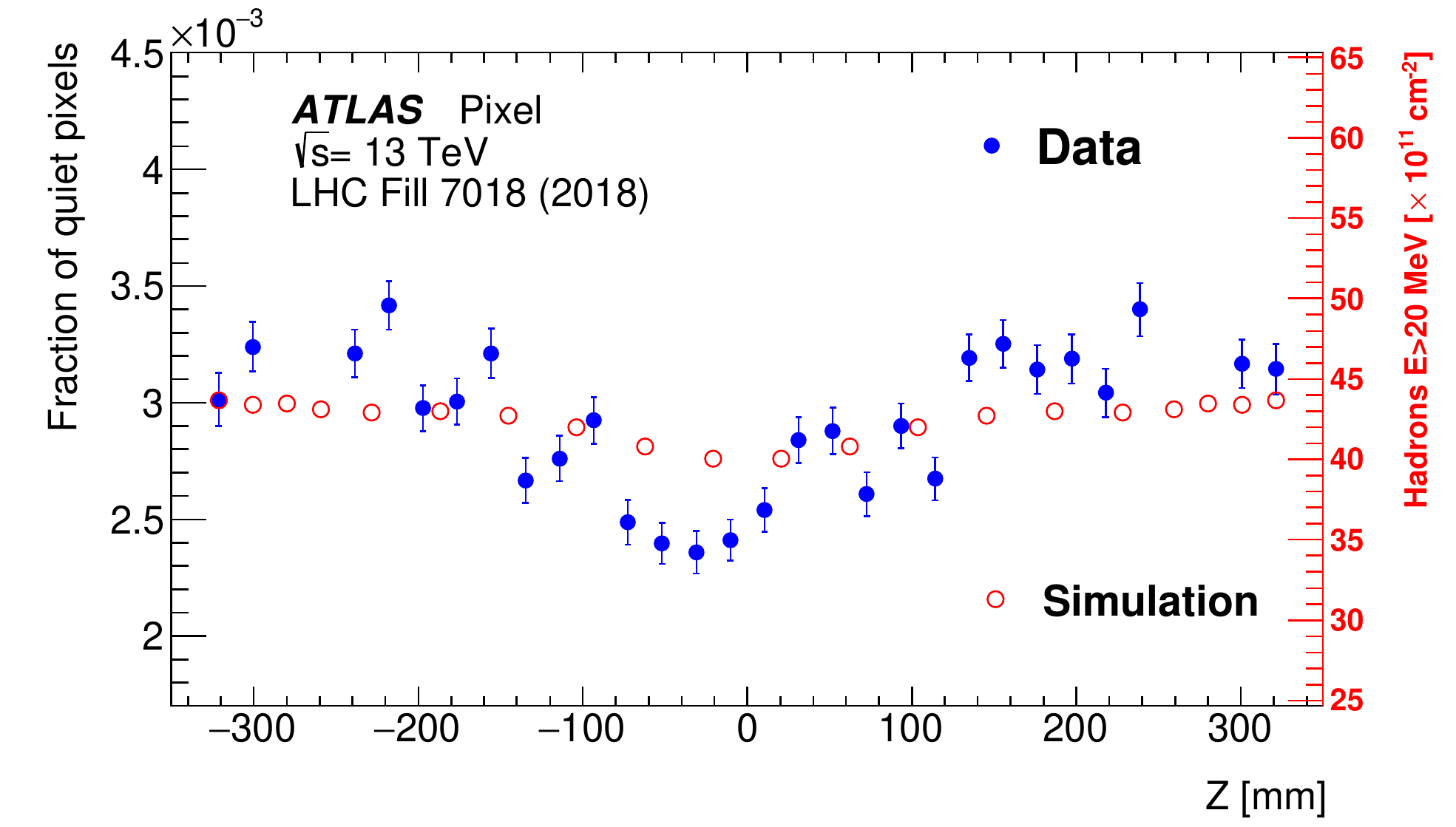}
\end{overpic}
\vspace{0.5cm}
\caption{Average fraction of quiet pixels in each IBL chip ring after $\sim$480\,pb$^{-1}$ of data taking in LHC fill 7018 of 2018, compared with PYTHIA/FLUKA simulations. Four points are missing due to the reconfiguration tests described in section~\ref{subsec:PRRecon}.}
\label{fig:zDep_fill5163}
\end{figure}

Figure~\ref{fig:zDep_fill5163} shows the average fraction of quiet pixels in each chip ring after $\sim$480\,pb$^{-1}$ of data taking in LHC fill 7018. It is compared with the PYTHIA/FLUKA simulation, which is normalized to the average fraction in data.

\subsubsection{Measurement of the change of enable bit fraction during an LHC fill}~\label{sec::LHCfill}
\begin{figure}[h]
\centering
 \begin{overpic}[width=0.82\linewidth]{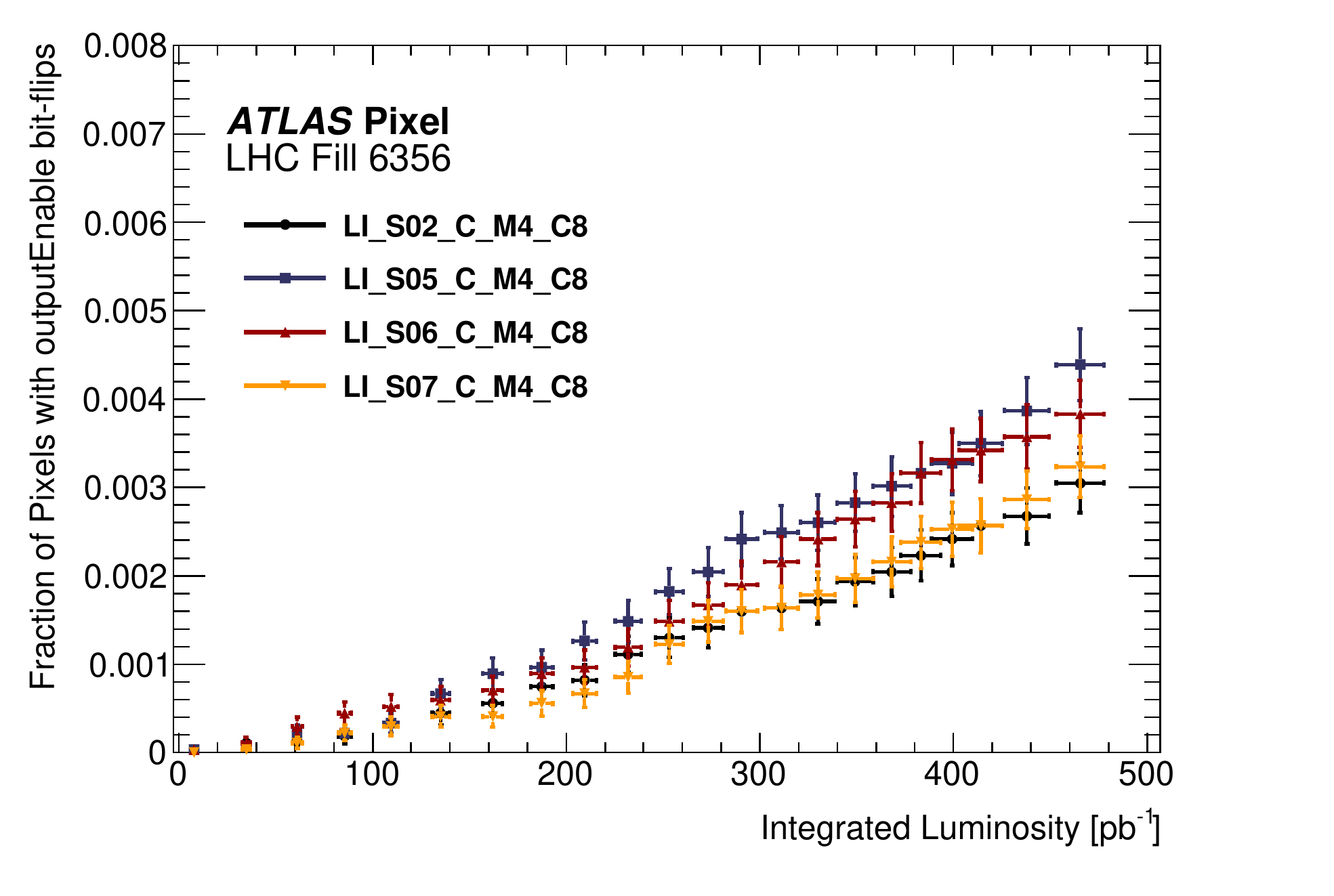}
\end{overpic}
\caption{Fraction of pixels in which enable bit $0\rightarrow1$ transition happens,  as a function of the integrated luminosity.  Four different FE-I4B chips placed on various IBL 3D modules are studied in LHC fill 6356. The numbering of the FE-I4B chips on 3D modules is in 
Ref.~\cite{ref:ATL-INDET-PUB-2014-006} and in the Appendix.}
\label{fig:enable-fill}
\end{figure}
A measurement of the change in the enable bit was carried out by disabling the output enable bits of all pixels in four FE-I4B chips at the beginning of LHC fill 6356. The pixels firing at the passage of a charged particle indicate SEU-induced transitions of output enable bits.
Figure~\ref{fig:enable-fill} shows the fraction of pixels, initially disabled at the beginning of the LHC fill, that are enabled by SEU. The rate of increase in enable bit $0\rightarrow 1$ transitions obtained during this LHC fill is consistent with the increase rate of the quiet pixels in figure~\ref{fig:fill5163-2016b}.

\subsubsection{Broken primary clusters}~\label{sec:broken-clusters}
The number of clusters produced by primary charged particles from a 13 TeV
$pp$ collision vertex
(referred to as ``primary clusters") is proportional to luminosity~\cite{ref:PCC}. Quiet pixels can cause clusters to be broken. In the 3D modules, the average length of a primary cluster is nine pixels, so broken clusters are a very sensitive probe of quiet pixels caused by a flip of the SEU enable bit from $1\rightarrow 0$. 

\begin{figure}[h]
\centering
 \begin{overpic}[width=0.49\linewidth]{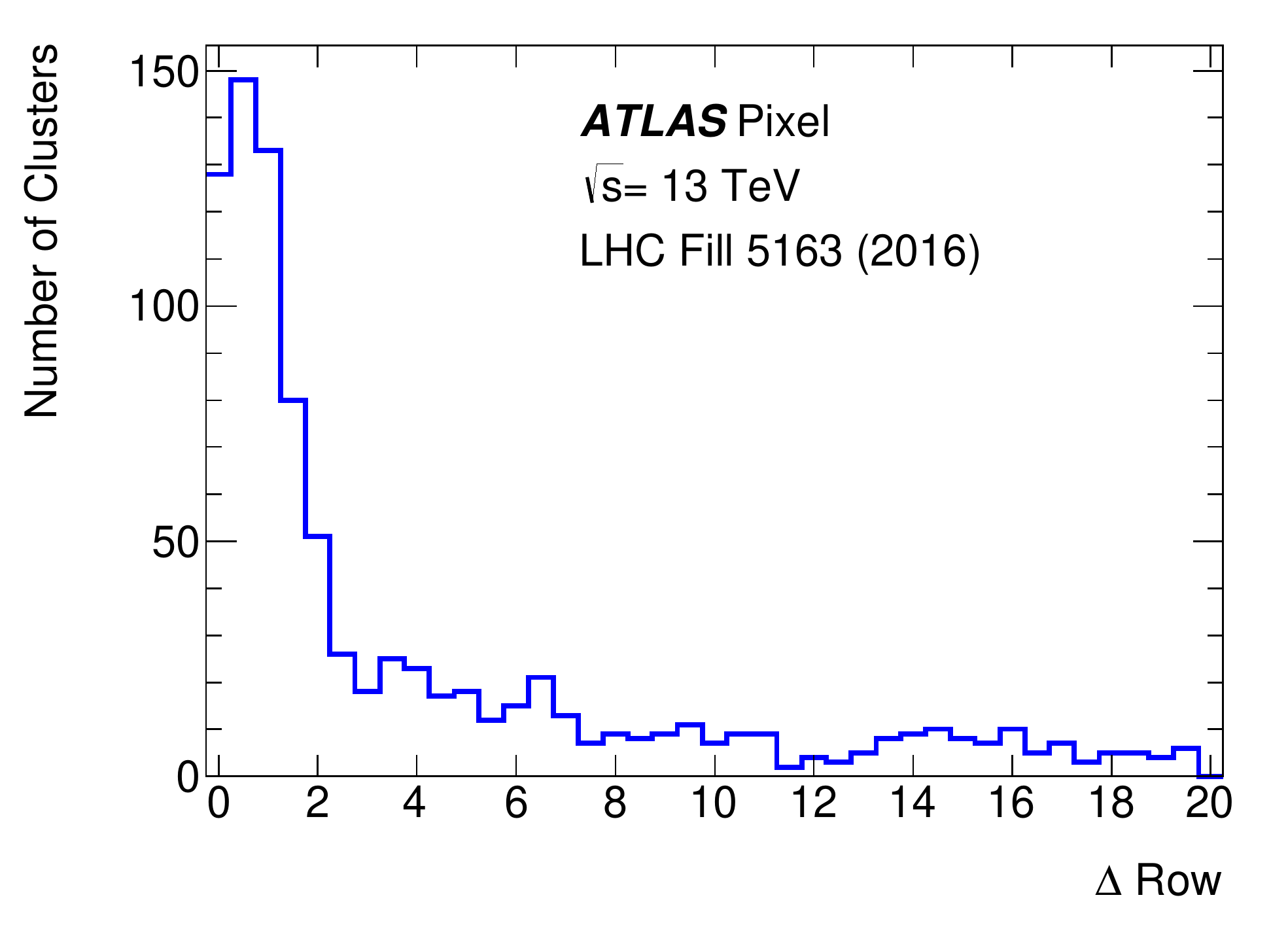}
\put(50,-5){(a)}
\end{overpic}
 \begin{overpic}[width=0.49\linewidth]{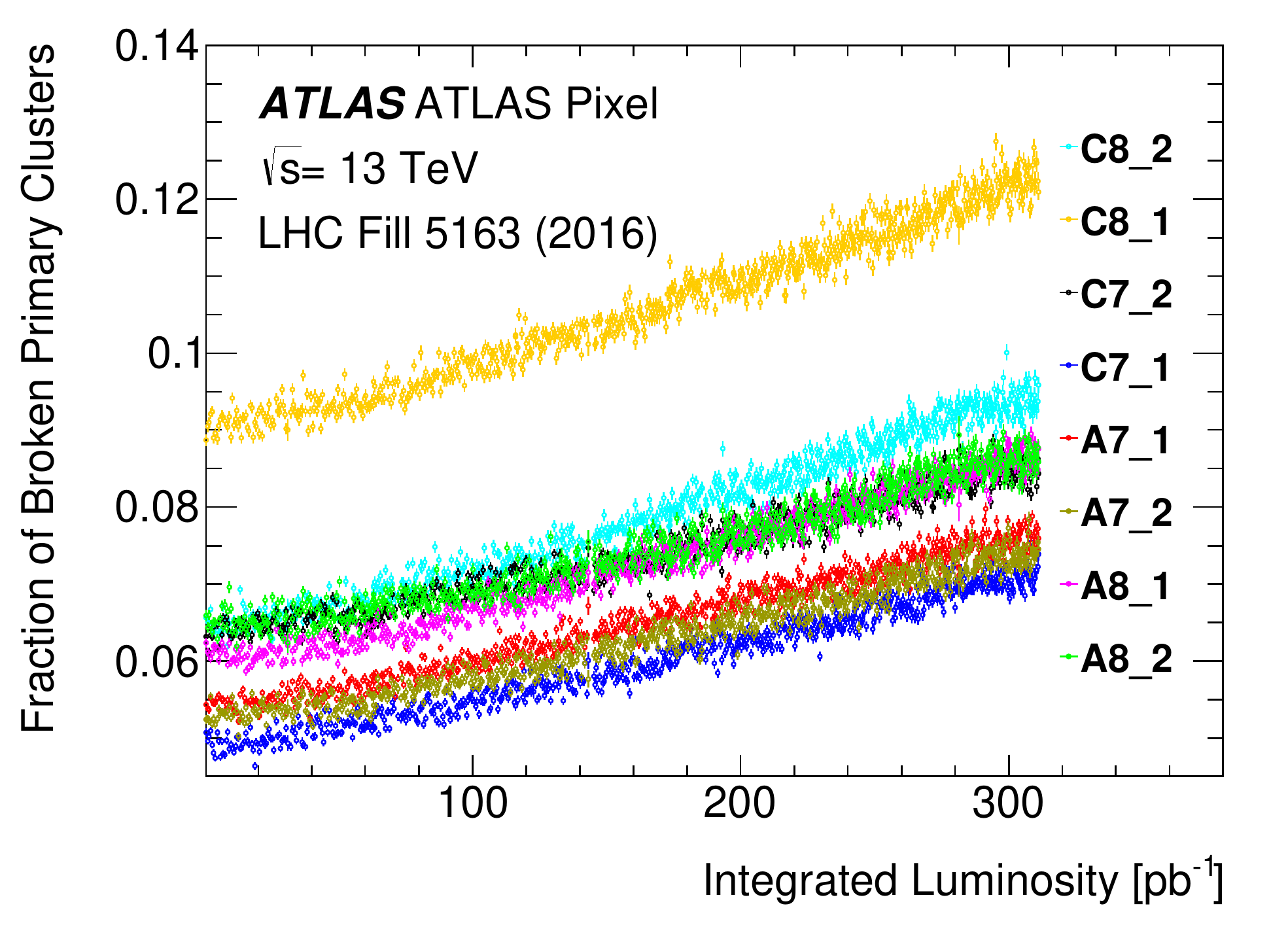}
\put(50,-5){(b)}
\end{overpic}
\vspace{0.5cm}
\caption{(a) $\Delta$row for all pairs of clusters separated by a one-pixel gap along the $z$ direction. The data shown is from the most forward 3D module on the negative-$z$ side, taken during the first LB of fill 5163 in 2016. (b) Fraction of broken primary clusters versus integrated luminosity in the same fill, in the 8 rings of 3D modules. }
\label{fig:broken5163}
\end{figure}

Figure~\ref{fig:broken5163}(a) shows the center-to-center distance along the azimuthal direction ($\Delta$row) for all pairs of clusters separated by a one-pixel gap along the $z$ direction. The data shown is from the most forward 3D module on the negative-$z$ side, taken during the first Luminosity Block (LB)
of fill 5163 in 2016. If two clusters are separated by a one-pixel gap along $z$ and $\Delta$row is no larger than three pixels, then these clusters are assumed to be broken from one long cluster. Background where two clusters are close together due to random coincidence is flat in $\Delta$row. This background level depends on the cluster density. By requiring that the longitudinal length of the merged cluster be consistent with a primary particle, broken background clusters from sources other than primary particles can be excluded from the $\Delta$row distribution ~\cite{ref:PCC}. The number of broken primary clusters in each LB is obtained by fitting the $\Delta$row distribution with a double-Gaussian describing the signal and a first-order polynomial describing the background. The total number of primary clusters in each LB is obtained by fitting the longitudinal cluster length distribution in Ref.~\cite{ref:PCC}.

Figure~\ref{fig:broken5163}(b) shows the fraction of broken primary clusters versus integrated luminosity in fill 5163 from 2016, in the 8 rings of 3D modules. The 14 modules in the same ring have been studied together to achieve higher statistics.  Module-to-module differences in the initial number of broken clusters indicate different fractions of silent pixels in the different rings.
All of the rings show a similar increase in the number of broken clusters with luminosity.

\subsubsection{Noisy pixels}~\label{sec:noisy}
Noisy pixels are studied with the same LHC fill as above, and the analysis strategy is similar to that used to study quiet pixels.
A pixel is defined to be noisy if it fired more than 350 times during the \SI{16}{\per\pico\barn} slice of data taking.

\begin{figure}[h]
\centering
 \begin{overpic}[width=0.82\linewidth]{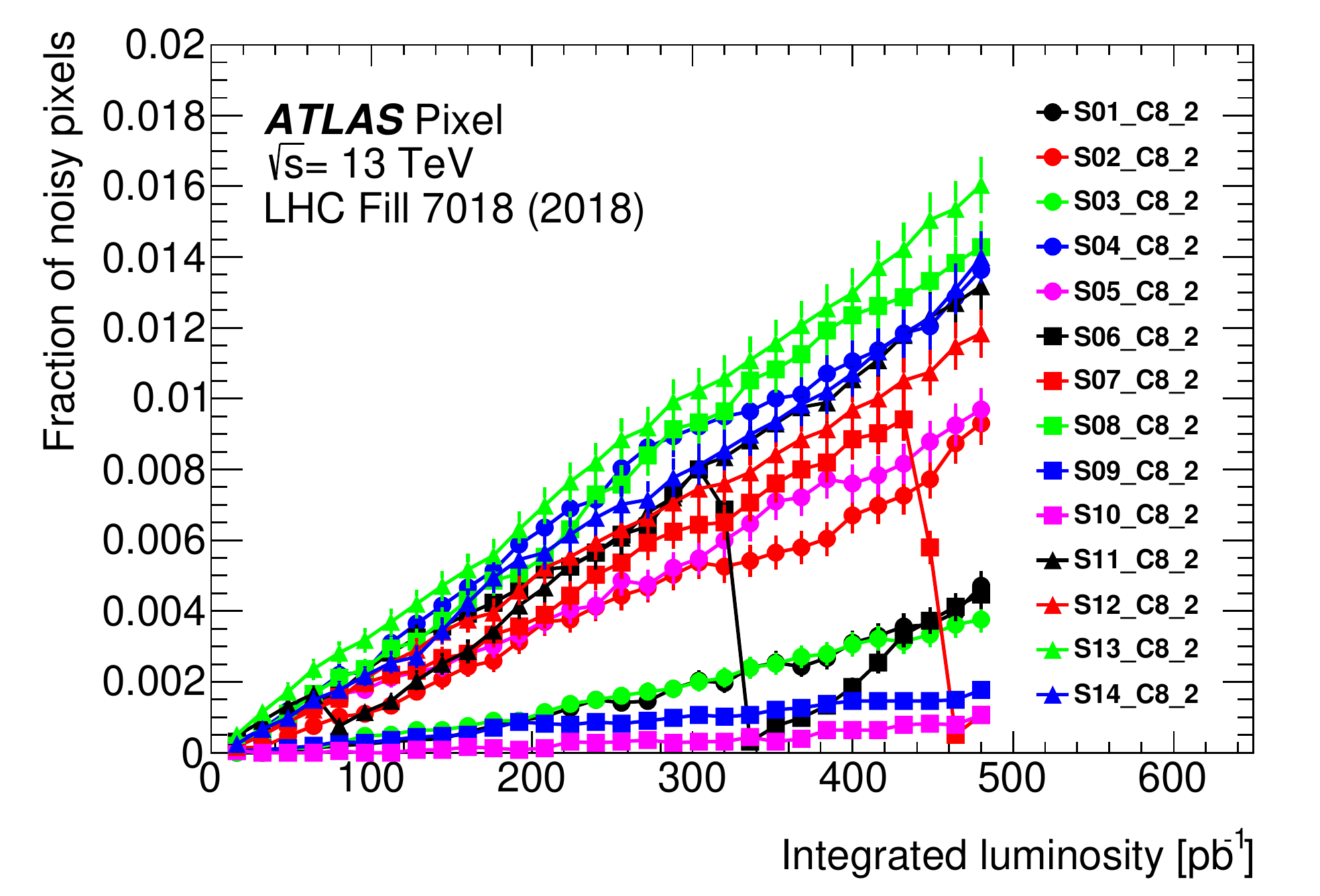}
\end{overpic}
\vspace{0.15cm}
\caption{
Fraction of noisy pixels versus integrated luminosity in the fill 7018, shown in all 14 modules in the most forward IBL ring on the negative-$z$ side. The 14 modules are displaced clockwise around the beam pipe with the first one at $\phi$=\ang{-6.1}. Long pixels and region of L-shape noise were removed from the analysis. }
\label{fig:fill5163-2016c}
\end{figure}

Figure~\ref{fig:fill5163-2016c} shows the fraction of noisy pixels versus integrated luminosity in the 14 modules in the most forward IBL ring on the negative-$z$ side. 
 We assume the pixels become noisy due to a threshold decrease.
In two modules one can observe a fast drop in the number of noisy pixels
due to the reconfiguration of these modules.
 How SEU effects make pixels noisy is explored in section~\ref{sec::TDAC-MSB}.

\subsection{Noise due to TDAC MSB bit flip}~\label{sec::TDAC-MSB}
The threshold in each pixel is determined by:
\vspace{-0.5cm}
\begin{center}
\begin{equation}
\label{eq:threshold}
\textrm{Threshold}=f_1(\textrm{VthinAlt\_Coarse})+f_2(\textrm{VthinAlt\_Fine})+f_3(\textrm{TDAC}\cdot \textrm{TdacVbp})
\end{equation}
\end{center}
VthinAlt\_Coarse and VthinAlt\_Fine are two 8-bit global registers for coarse and fine adjustment of the global threshold. TdacVbp (8-bit global register) sets the step size of TDAC.  The bits of the global registers are not easily flipped, thanks to the use of triply redundant latches.  This analysis considers 
noisy pixels resulting from the bit flip of TDAC.

\begin{figure}[h]
\centering
 \begin{overpic}[width=0.82\linewidth]{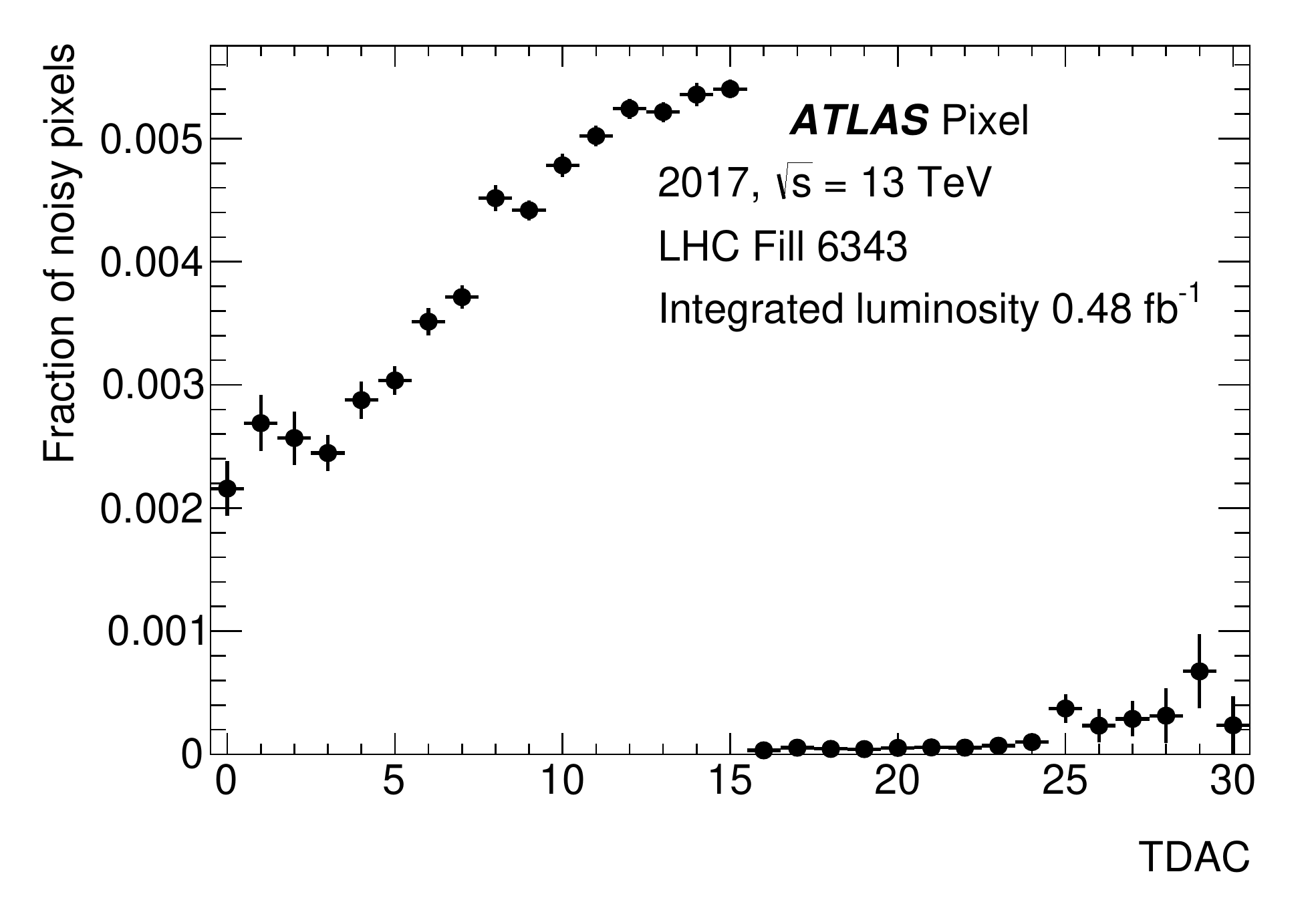}
\end{overpic}
\caption{Fraction of noisy pixels as a function of pixel TDAC during empty bunches of LHC fill 6343. TDAC values for each pixel are taken from the initial pixel configuration. The pixels with more than 200 hits in this fill are treated as noisy.}
\label{fig:noiseVSTDAC}
\end{figure}

Figure~\ref{fig:noiseVSTDAC} shows the fraction of noisy pixels as a function of pixel TDAC during empty bunches of a typical LHC fill (6343 in 2017). TDAC values for each pixel are taken from the initial pixel configuration. The pixels with more than 200 hits in this fill are defined as noisy. Low values of TDAC correspond to high thresholds. The increase in the fraction of noisy pixels with initial TDAC $< 15$ indicates that some pixels become noisy due to the SEU flip $0\rightarrow1$ of the most significant bit (MSB) of TDAC, which lowers the pixel threshold by $\sim$\SI{1850}{\e}. No correlation of the noise with FDAC values was observed.

\begin{figure}[h]
\centering
 \begin{overpic}[width=0.52\linewidth]{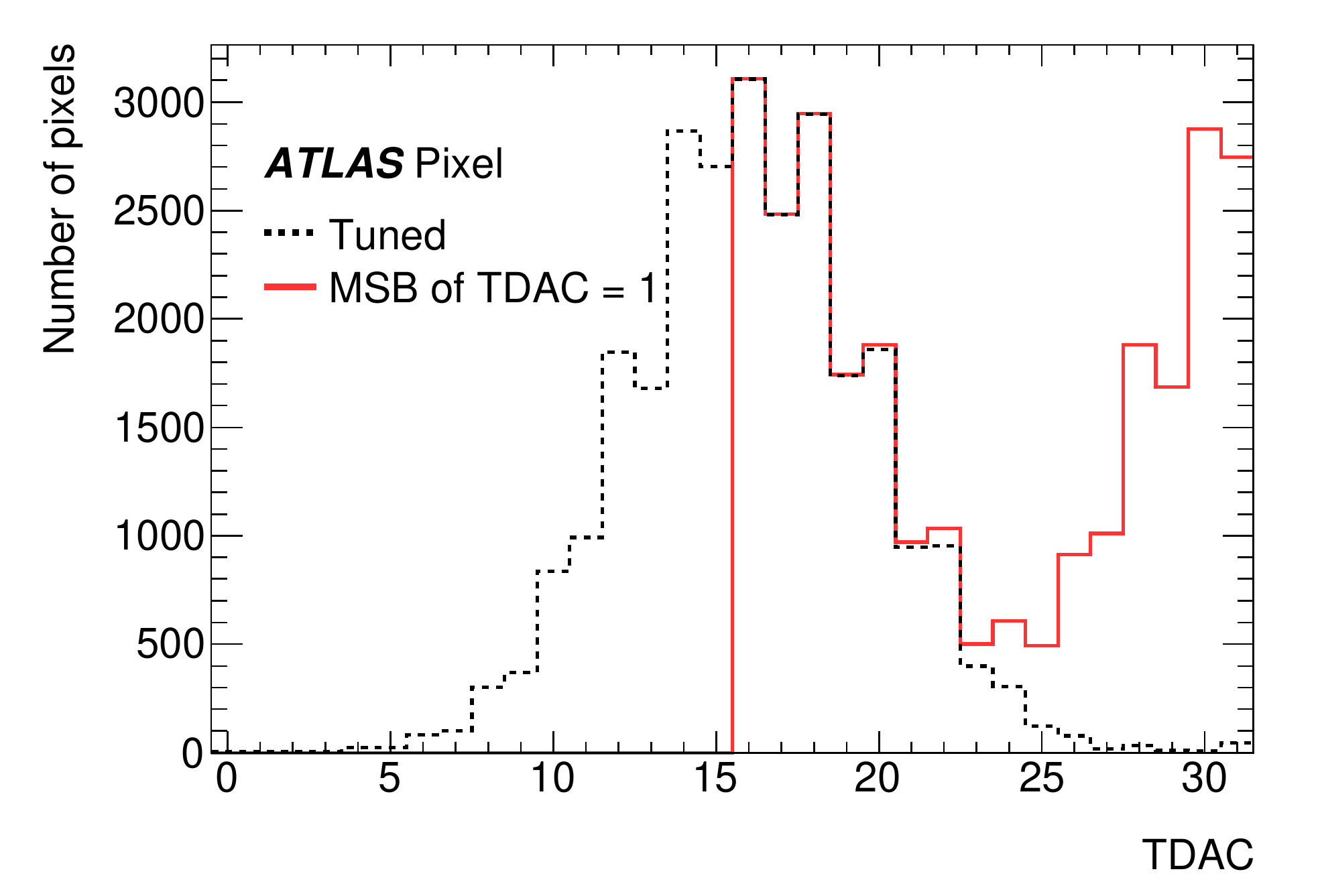}
\put(1,5){(a)}
\end{overpic}
 \begin{overpic}[width=0.49\linewidth]{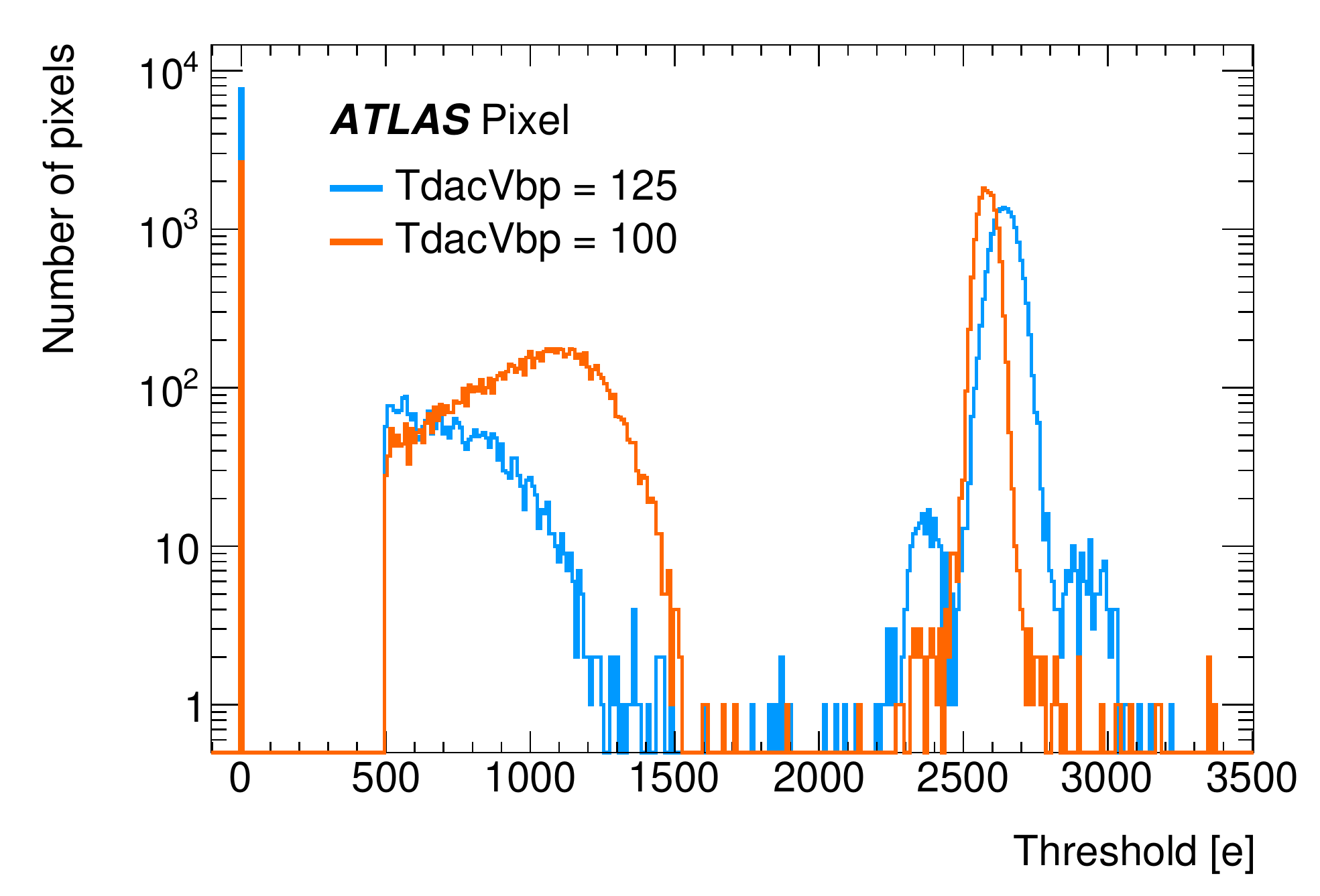}
\put(50,-5){(b)}
\end{overpic}
 \begin{overpic}[width=0.49\linewidth]{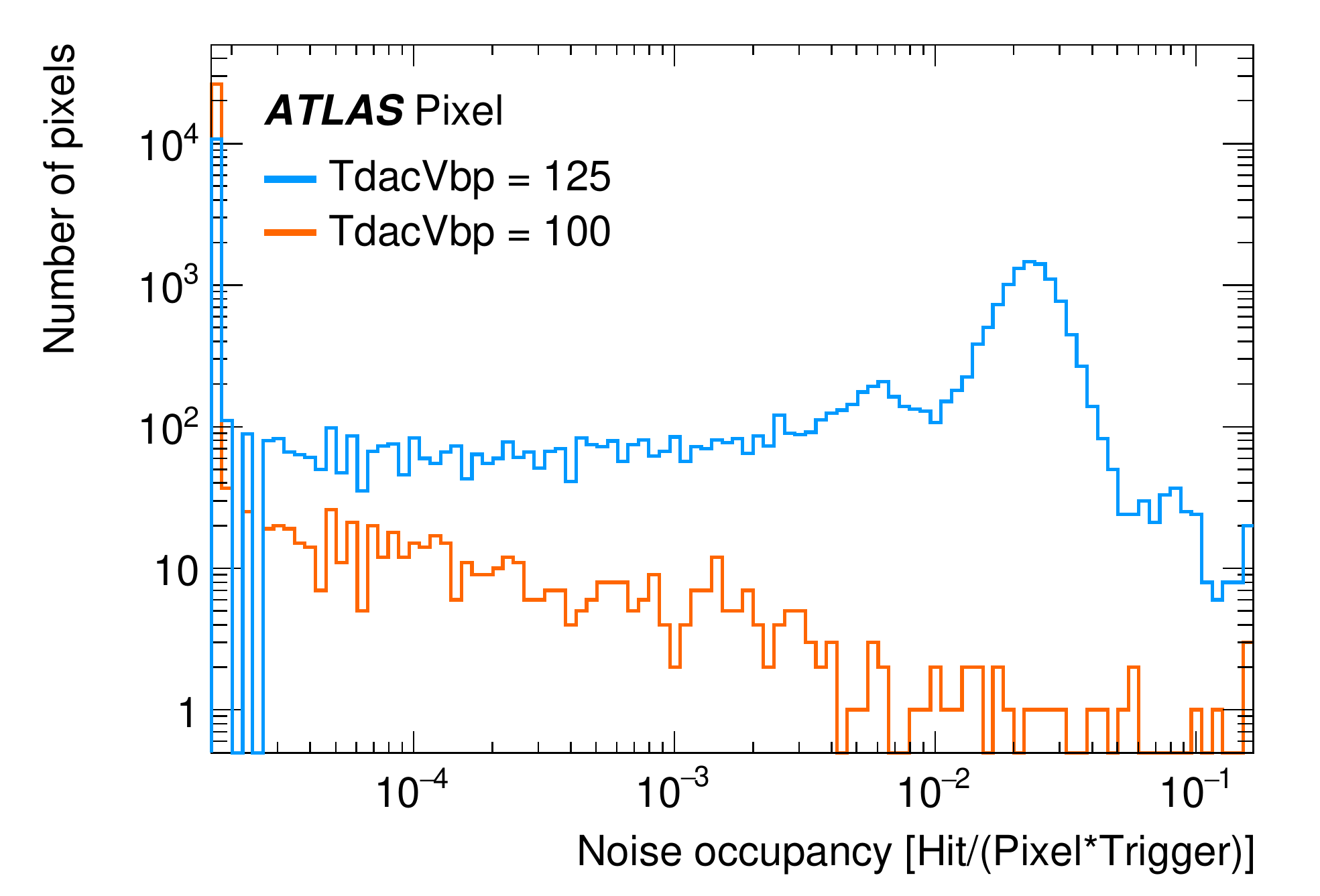}
\put(50,-5){(c)}
\end{overpic}
\vspace{0.15cm}
\caption{(a) TDAC distribution when TdacVbp is set to 100 and the threshold is tuned to 
\SI{2500}{electrons}
 in an FE-I4B chip (dash). To mimic an SEU on the MSB of the TDAC, the MSB is set to 1 (solid). The chip is bonded to a non-irradiated sensor. (b) Threshold distributions after the MSB shift to 1 and the module was retuned with a different TdacVbp=125. (c) Noise occupancies of each pixel after the MSB shift. The last bin includes overflow pixels. }
\label{fig:testing}
\end{figure}

\subsection{Manual reproduction of the TDAC bit flip}~\label{sec::TDAC-mimic}
To mimic an SEU on the MSB of TDAC, the MSB is set to 1 in the laboratory for all pixels on an individual FE-I4B test chip, which has been tuned with a target threshold of \SI{2500}{\e}. The chip is bonded to a non-irradiated sensor.
TdacVbp (step size of the least significant bit of the in-pixel threshold tuning DAC) is set to 100. 
As shown in figure~\ref{fig:testing}(a), the TDAC distribution is shifted when TDAC MSB is set to 1 for all pixels. 

Figure~\ref{fig:testing}(b) shows the threshold distributions after the MSB of TDAC in each pixel is set to 1 and the module was retuned with TdacVbp=125.
Thresholds below \SI{500}{\e} cannot be measured (due to a lower limit of the charge injection) and are defined to be 0. The threshold of some pixels decreases due to a flip of TDAC-MSB. 

The average noise occupancy distribution in figure~\ref{fig:testing} (c) is obtained with a noise scan
\footnote[3]{Noise scan consists of reading out all pixels in the matrix without any charge injection.}
 by sending $3\times10^7$ triggers. Many pixels become noisy due to the threshold decrease caused by the $0\rightarrow1$ transition of TDAC-MSB. The FE-I4B module is more robust against SEUs when TdacVbp is smaller, but some pixels cannot be tuned at low TdacVbp.

The fraction of noisy pixels is different among different chips.
Could these differences be explained due to the different TdacVbp values?
In fact 99\% of FE-I4B chips in the IBL have TdacVbp=180. In particular, 13 out of 14 chips included in figure~\ref{fig:fill5163-2016c} have TdacVbp=180 (the module denoted by black circles has TdacVbp=140). Therefore, differences in the fraction of noisy pixels cannot be explained by different values of TdacVbp. These differences are partly due to the differences in the mean TDAC values in the chip and probably also related to different offsets of the effective global thresholds between the modules or to differences in the global noise levels.

\subsection{L-shaped noise map in 3D sensor modules}~\label{sec::Lshape}
\begin{figure}[h]
\centering
 \begin{overpic}[width=1.00\linewidth]{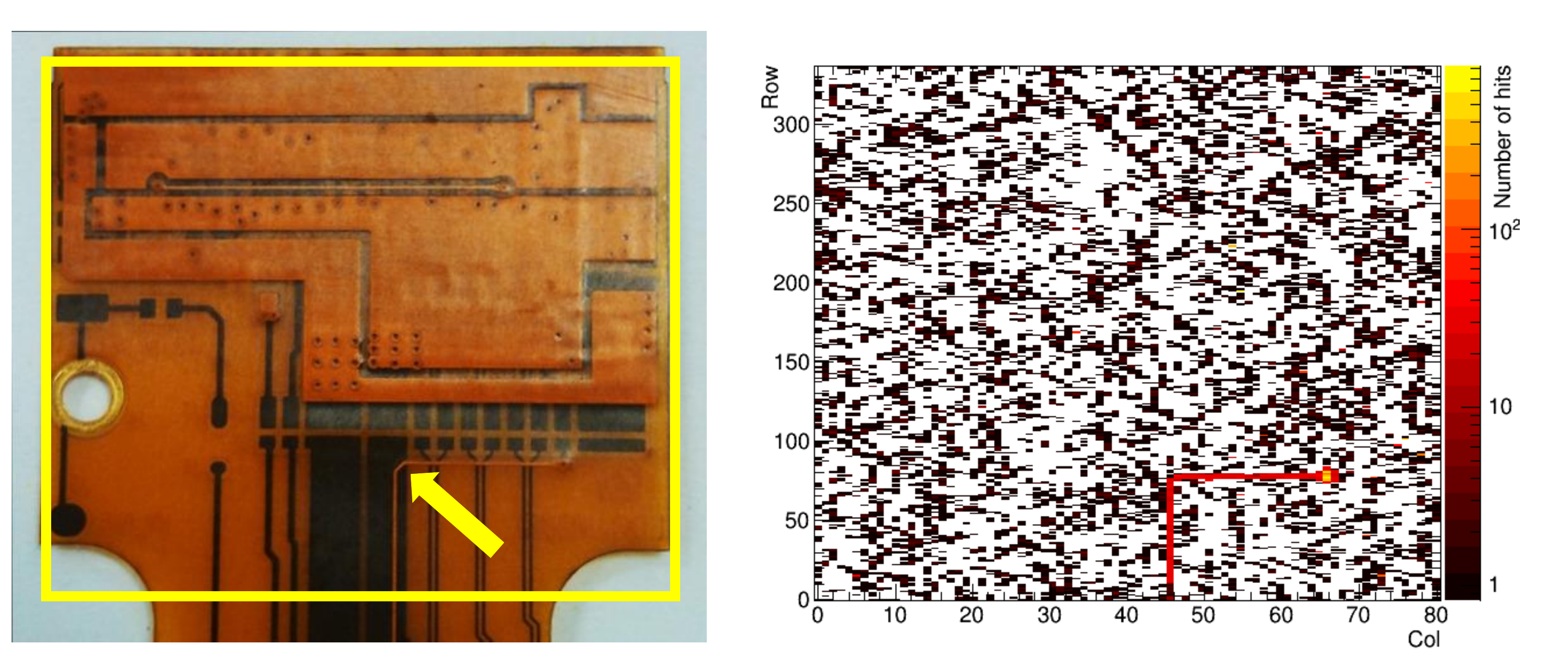}
\put(25,-5){(a)}
\put(75,-5){(b)}
\end{overpic}
\vspace{0.5cm}
\caption{(a) Image of the flex cable with L-shaped HitOr line attached
to the sensor. (b) The map (columns versus rows) of L-shaped noise in the reworked FBK-type 
 3D IBL module LI-S14-C-M4-C8-2 accumulated by random triggers during empty bunches of LHC fill 6239 with integrated luminosity \SI{0.13}{\per\femto\barn}.}
\label{fig:Lshape}
\end{figure}
One consequence of SEE effects in FE-I4B was the 
observation of an L-shaped noise structure in reworked ($\sim$50\% of all) FBK-type
\footnote[4]{Modules with 3D sensors produced at FBK factory~\cite{ref:IBL}.}
 3D sensor IBL
modules (See figure~\ref{fig:Lshape}). On these 3D modules, the output of the HitOr signal was accidentally wire bonded
to the flex cable attached to the back side of the sensor. On the FBK sensors, the metallization on the back side
is not continuous, but has a grid-like structure,
which allows some cross talk from the HitOr line on the flex cable into the 3D
sensor. Normally the HitBus outputs are disabled, so during laboratory and
low luminosity runs this cross talk was not observed. But during high
luminosity runs, several HitBus bit flips are produced by SEE effects, and the
signals on the HitOr flex cable line induce cross talk signals in the
pixels under this line with this particular L-shape. This effect has	
no visible consequence for physics, because the 3D modules are outside the tracking acceptance  of $|\eta|$ < 2.5, but it provides several lessons for future	
detectors: to use full backside metallization in the sensor, to	
avoid close contact of flex cables with the sensors, and to remove	
unused wire bonds between the chip and the cables.

%% file: documents/readback.tex
The FE-I4B provides the functionality to read back the content of the DICE latches by copying the latch content in the Shift Register (SR) for each double column and transmitting it back to the Readout Drivers (RODs).
The SEU measurement scheme is to write to a register in the device being tested, let a certain fluence of particles traverse the device, and then read back the register.
\subsection{Single pixel configuration read-back measurements ~\label{sec:pixelreadback}}
The fraction of pixels in which the bit state flips ($\frac{N_i(0) - N_i}{N_i}$ ($i=0,1$)) after taking data of integrated luminosity $\mathcal{L}$ depends on the probability of $0\rightarrow1$ ($P_{0\rightarrow1}$) and $1\rightarrow0$ ($P_{1\rightarrow0}$) transitions: 
\begin{align}
\label{eq:NN-1}
\frac{N_1(0) - N_1}{N_1} = P_{1\rightarrow0}\cdot \mathcal{L} \\
\frac{N_0(0) - N_0}{N_0} = P_{0\rightarrow1}\cdot \mathcal{L}
\end{align}

\begin{figure}[h]
\centering
 \begin{overpic}[width=0.49\linewidth]{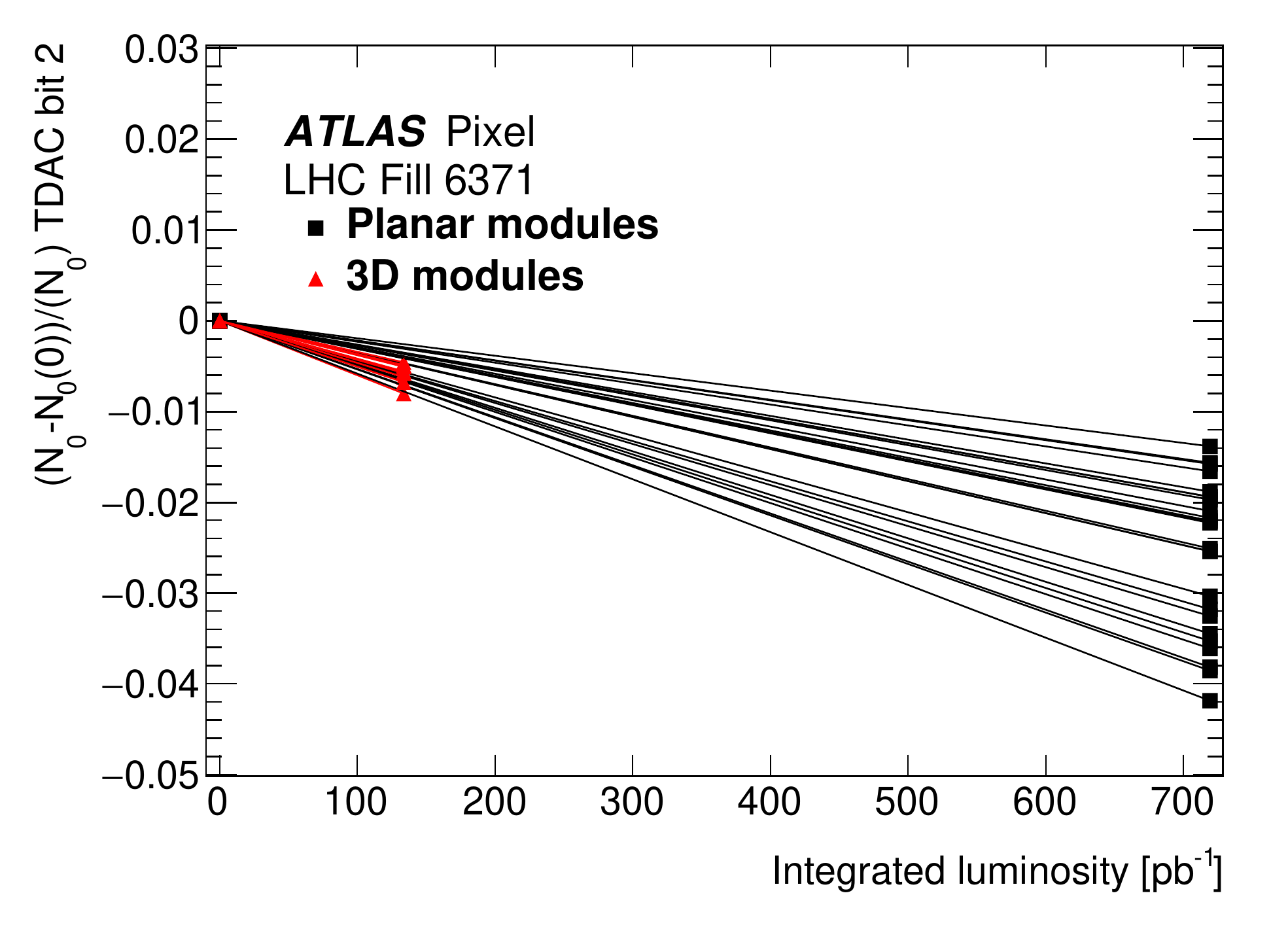}
\put(50,-5){(a)}
\end{overpic}
 \begin{overpic}[width=0.49\linewidth]{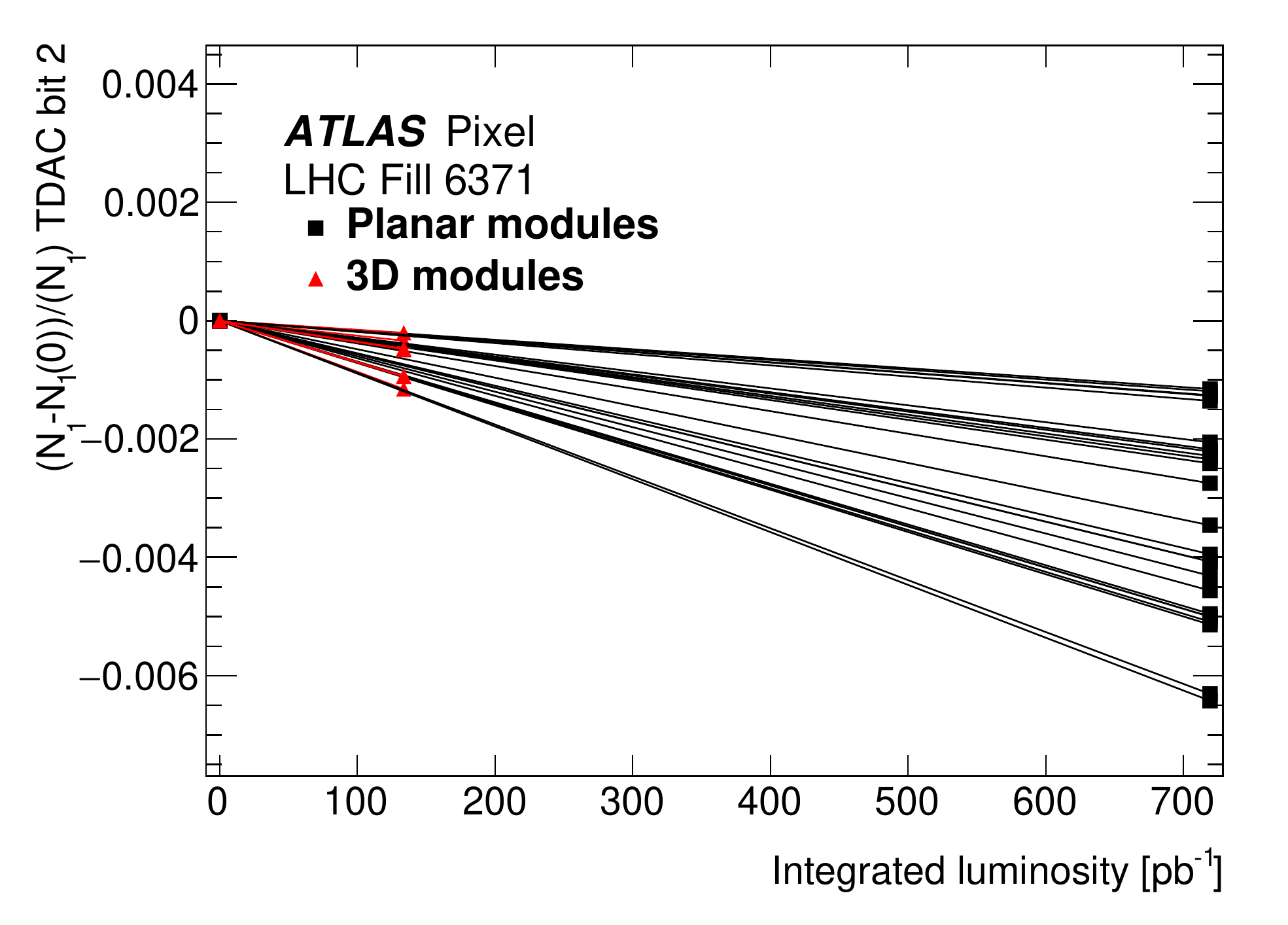}
\put(50,-5){(b)}
\end{overpic}
\vspace{0.5cm}
\caption{Fraction of pixels in which TDAC-bit2 flips, measured at the beginning of LHC fill 6371 and after about \SI{150}{\per\pico\barn} of integrated luminosity for 3D modules and after the end of LHC fill for planar modules. (a) $0\rightarrow1$. (b) $1\rightarrow0$.}
\label{fig:TDAC-bit2}
\end{figure}
In figure~\ref{fig:TDAC-bit2}, the 
status
 of TDAC-bit2 in each pixel is checked
at the beginning of LHC fill 6371 and after about \SI{150}{\per\pico\barn}
of data for 3D sensors and after the end of LHC fill for planar modules, respectively.  Read-back cannot be performed while a FE receives trigger signals, so there are only two read-backs for planar modules: before the start of the collisions and after beam dump. The intermediate point is from 3D modules.
The read-back for 3D modules were done more frequently, because 3D modules are
outside tracking acceptance of $|\eta|$<2.5, so for small number of runs
we could remove 3D information.
 The first point is zero by construction.
$P_{0\rightarrow1}$ and $P_{1\rightarrow0}$ are extracted from these measurements. This procedure allows for an independent measurement of the SEU probabilities for each of the 13 pixel latches.

\begin{figure}[h]
\centering
 \begin{overpic}[width=0.52\linewidth]{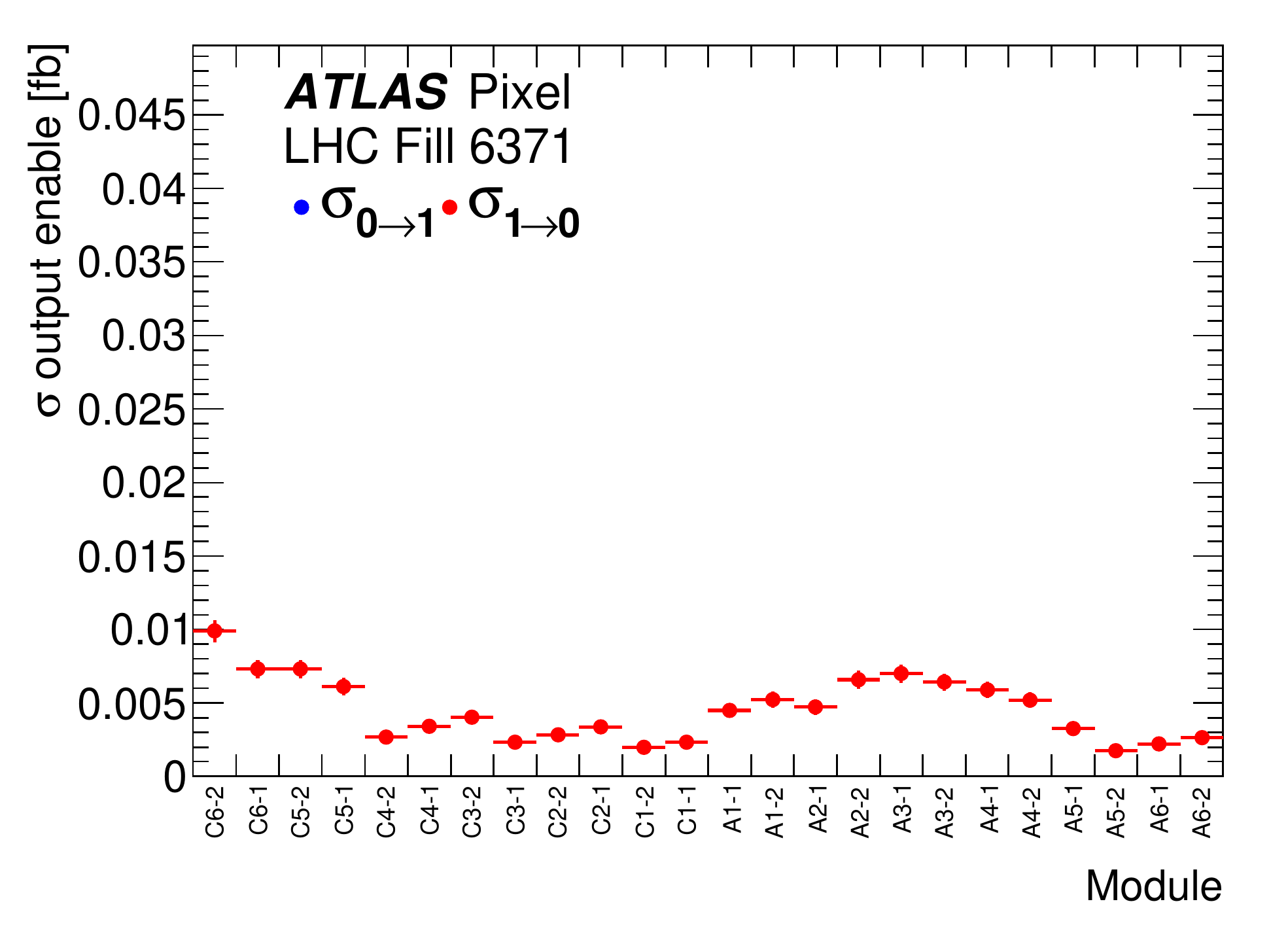}
\put(1,5){(a)}
\end{overpic}
 \begin{overpic}[width=0.49\linewidth]{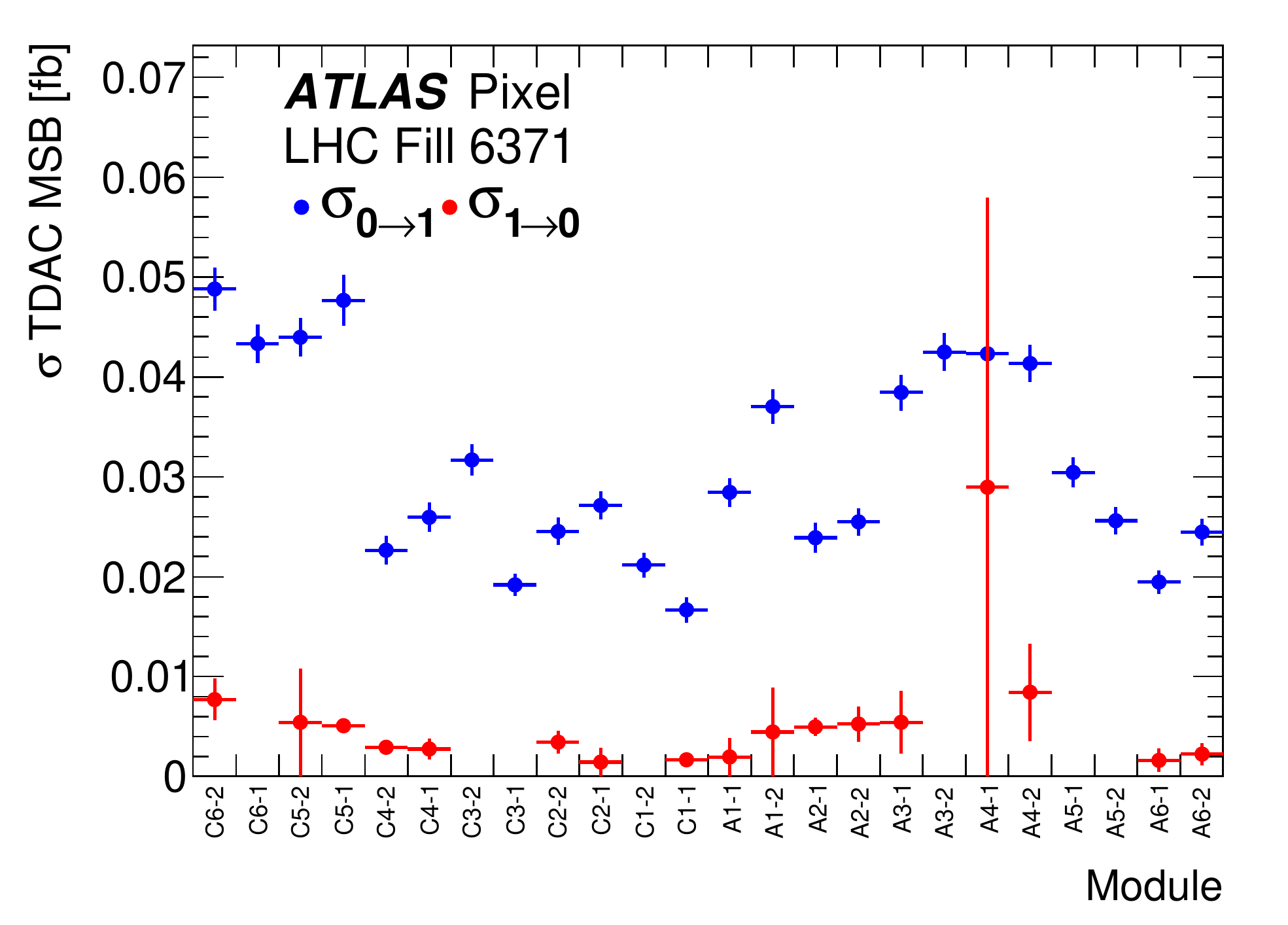}
\put(50,-5){(b)}
\end{overpic}
 \begin{overpic}[width=0.49\linewidth]{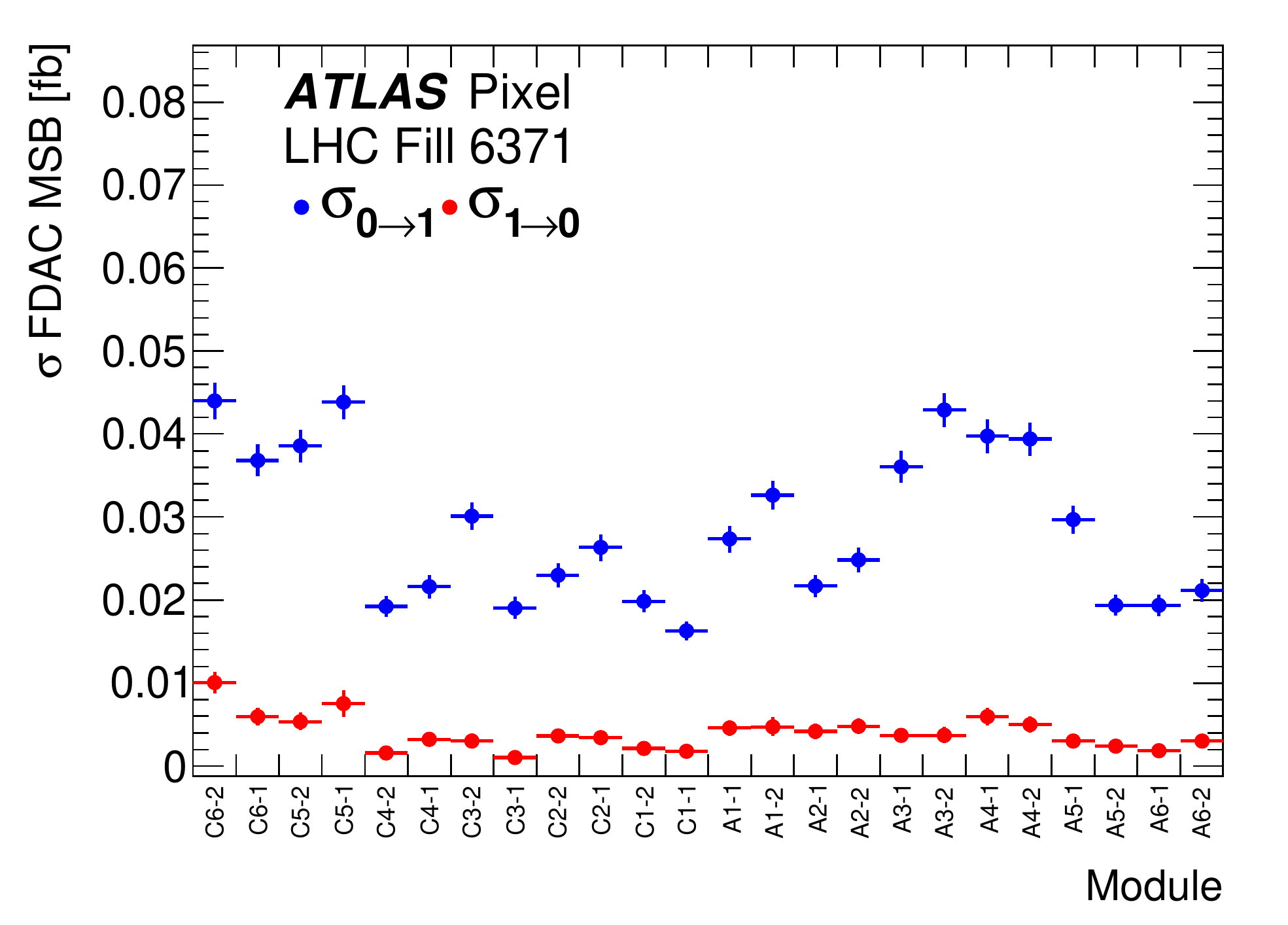}
\put(50,-5){(c)}
\end{overpic}
\vspace{0.15cm}
\caption{The probability for a latch to undergo a $0\rightarrow1$ or $1\rightarrow0$ transition measured for the output enable bit (a), TDAC-MSB (b), and FDAC-MSB (c) for FE-I4B in each planar module from negative-$z$ side to positive-$z$ side. The $P_{0\rightarrow1}$ is not present for the output enable bit because almost all pixels in FE-I4B have this bit set to 1, and a measurement of such transition is not possible for this latch.}
\label{fig:sgm-results}
\end{figure}

The $0\rightarrow1$ and $1\rightarrow0$ transition probabilities have been measured for the pixel latch memories. Only the results of the latches most important for DAQ operations are reported. In figure~\ref{fig:sgm-results} (a), the $P_{0\rightarrow1}$ of output enable bit has not been calculated because most of the pixels are enabled initially.
In figure~\ref{fig:sgm-results} (b) and (c), the $0\rightarrow1$ transition appears to be more frequent by a factor of $\sim5$ with respect to the $1\rightarrow0$ transition for the FEs investigated.
The simulation of the SEUs in DICE memories
~\cite{ref:AFTU}, ~\cite{ref:AFTUSimul},
~\cite{ref:SETSimul}
demonstrates that the dominant effect comes from glitches on the LOAD signal.
The LOAD signal stores into the DICE latch the current content of the SR. 
 In this case, the polarity of the transition depends
on the actual values stored in the SR. The last bit (out of 13 bits) loaded into the SR is the output enable bit. As the enable bit is usually 1, this favors $0\rightarrow1$ transitions.

\begin{figure}[h]
\centering
 \begin{overpic}[width=0.82\linewidth]{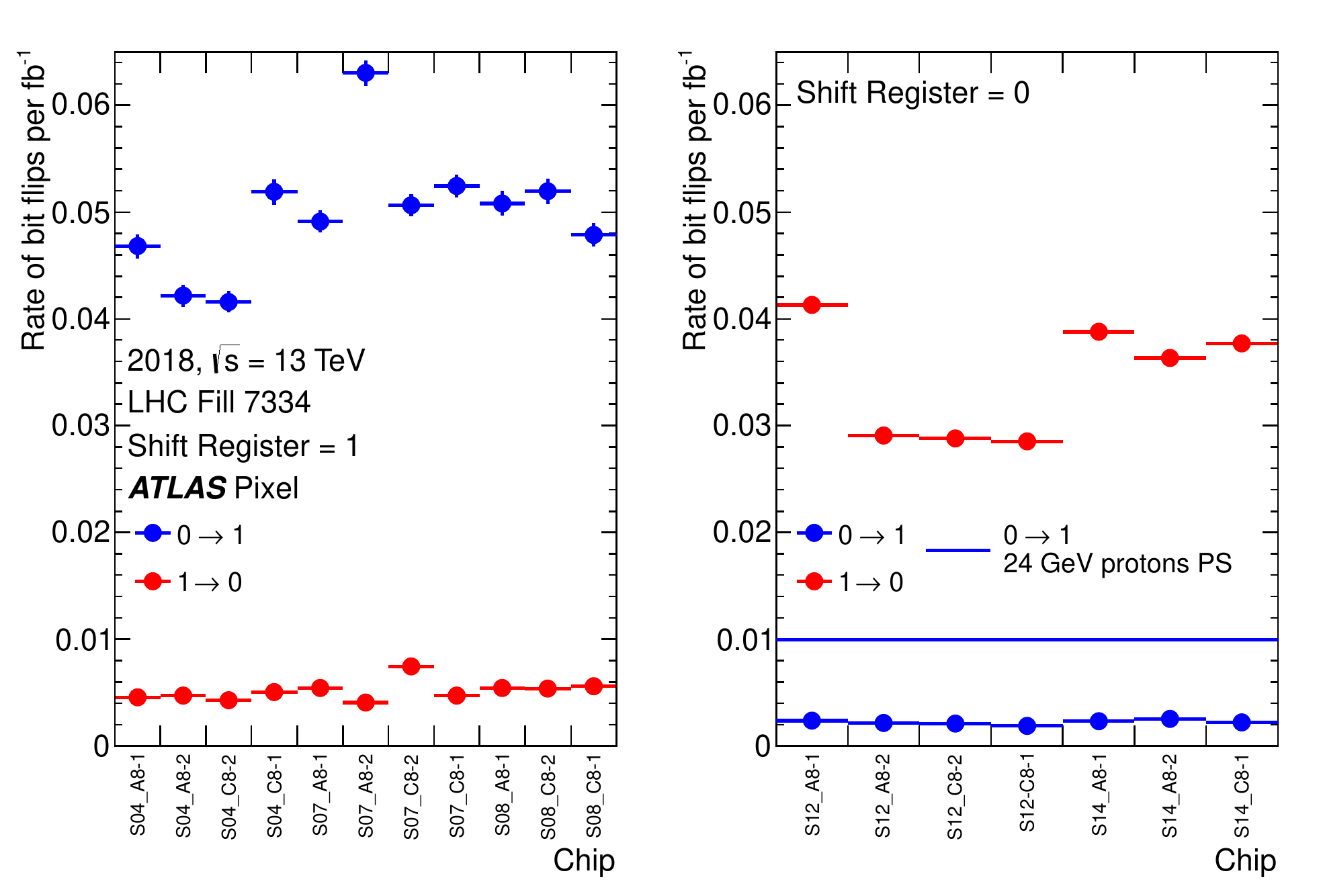}
\put(25,-1){(a)}
\put(75,-1){(b)}
\end{overpic}
\caption{
In figure (a), the Shift Register (SR) was set to 1, and $0\rightarrow1$ flips dominate due to the SET on the LOAD line, while low rate $1\rightarrow0$ flips are due to real memory SEU. In 
figure (b), the Shift Register was set to 0, and $1\rightarrow0$ dominate. The values of the Shift
Register are refreshed several times during the fill.
}
\label{fig:hrvschipsr10}
\end{figure}

These asymmetries were studied in detail during LHC fill 7334 with the read-back method.
The probability of a pixel memory bit flip strongly depends 
on the value stored in the input Shift Register. Pixel memory read-back results
 and chip simulations confirm that the FE-I4B SEEs are dominated by
 Single Event Transient effects (SET), which create
 fake LOAD signals. $0\rightarrow1$ flips in pixel memory are dominant when the SR is
 loaded with 1, and  $1\rightarrow0$  flips are dominant when SR is loaded with 0.
If the glitches on the LOAD line are created in the transistors close to
the memory cell, only one bit is reloaded, but if the glitches are created in the
 transistors upstream, several bits could be reloaded simultaneously. 
The average rate of the SEU/SET bit flips in pixel memory of FE-I4B per fb$^{-1}$ was studied in
 LHC fill 7334 as a function of the chip number. In figure~\ref{fig:hrvschipsr10} (a), the 
Shift Register (SR) was set to 1, and $0\rightarrow1$ flips dominate due to the SET on the LOAD line, while low rate $1\rightarrow0$ flips are due to real memory
 SEU. In figure~\ref{fig:hrvschipsr10} (b), the Shift Register was set to 0, and $1\rightarrow0$  flips 
dominate. The values of the Shift Register are refreshed several times during the
LHC fill. 
The extrapolation of the measurement of the SEU rate with 24 GeV protons at CERN PS is 
shown with a blue line in figure~\ref{fig:hrvschipsr10} (b). During the CERN PS measurement,
 the value of the SR was not refreshed, which may explain the higher rate of bit flips
 due to SET contributions. 
Figures~\ref{fig:hrvschipsr10} (a) and (b) present different chips at slightly different locations, so part of the
 difference may come from chip-to-chip process variations, tuning, and particle flux differences.

Simulations of SET effects in DICE memories
~\cite{ref:AFTU}, ~\cite{ref:AFTUSimul},
~\cite{ref:SETSimul} demonstrate
that most of the SET on the LOAD line are produced by heavy ionizing particles 
simultaneously crossing the pair of transistors T1 and T2 (or alternatively T3 and T4). 
Unfortunately, in the current implementation of the DICE latch (see figure~\ref{fig:dice_scematics})
 in FE-I4B, the transistor T1 is close to T2, and transistor T3 is close to T4.  Correlated glitches 
created by the same particle in both transistors are therefore relatively probable.
In the same simulations, it was shown that one can considerably reduce these glitches by
increasing the distance between the transistors T1-T2 and T3-T4.

\begin{figure}[h]
\centering
 \begin{overpic}[width=0.82\linewidth]{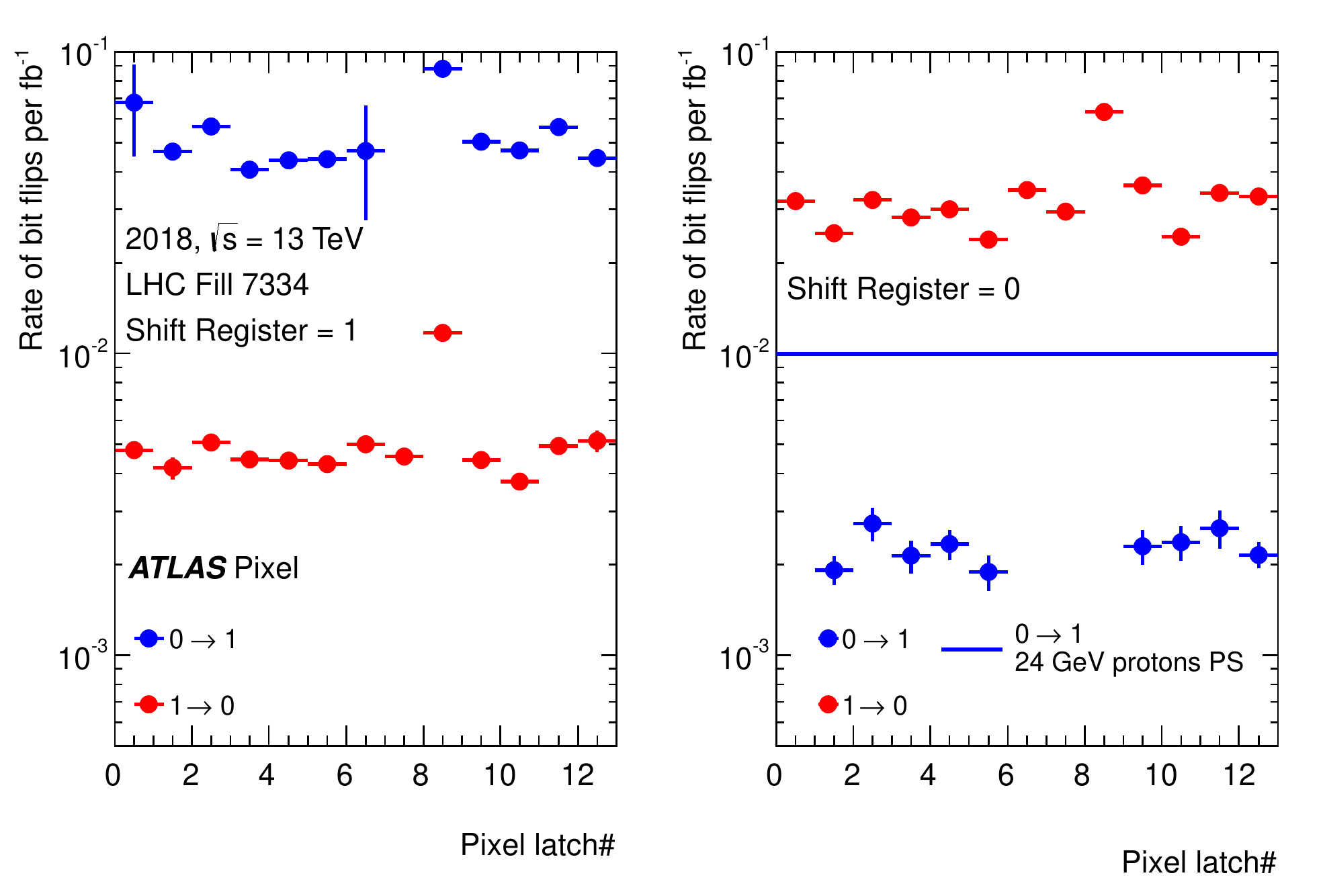}
\put(25,-1){(a)}
\put(75,-1){(b)}
\end{overpic}
\caption{
Average rate of SEU/SET bit flips in pixel memory of FE-I4B per fb$^{-1}$
in LHC fill 7334, as a function of bit number (0-12). In figure (a), the Shift
Register was set to 1, and $0\rightarrow1$ flips dominate due to the glitches on the LOAD line, while low rate $1\rightarrow0$ flips are due to real memory 
SEU. In figure (b), the Shift Register was set to 0, and $1\rightarrow0$ flips dominate.
The extrapolation of the measurement of the SEU rate with 24 GeV protons at CERN PS is 
shown with a blue line. During the CERN PS measurement,
 the value of the SR was not refreshed, which may explain higher rate of bit flips due to SET contributions. 
}
\label{fig:hrvsbitsr10}
\end{figure}
In the same fill, the average rate of the SEU/SET bit flips in pixel memory was studied
as a function of the bit number (0-12).
In figure~\ref{fig:hrvsbitsr10} (a), the
 Shift Register was set to 1, and $0\rightarrow1$ flips dominate due to the SET
 on the LOAD line, while low rate $1\rightarrow0$ flips are due to real memory
 SEU. In figure~\ref{fig:hrvsbitsr10} (b), the Shift Register is set to 0, and $1\rightarrow0$ flips
 dominate.
Flips in bit-8 (HitBus) are twice as frequent,
probably due to specific layout of that memory bit.

\begin{figure}[h]
\centering
\begin{overpic}[width=0.82\linewidth]{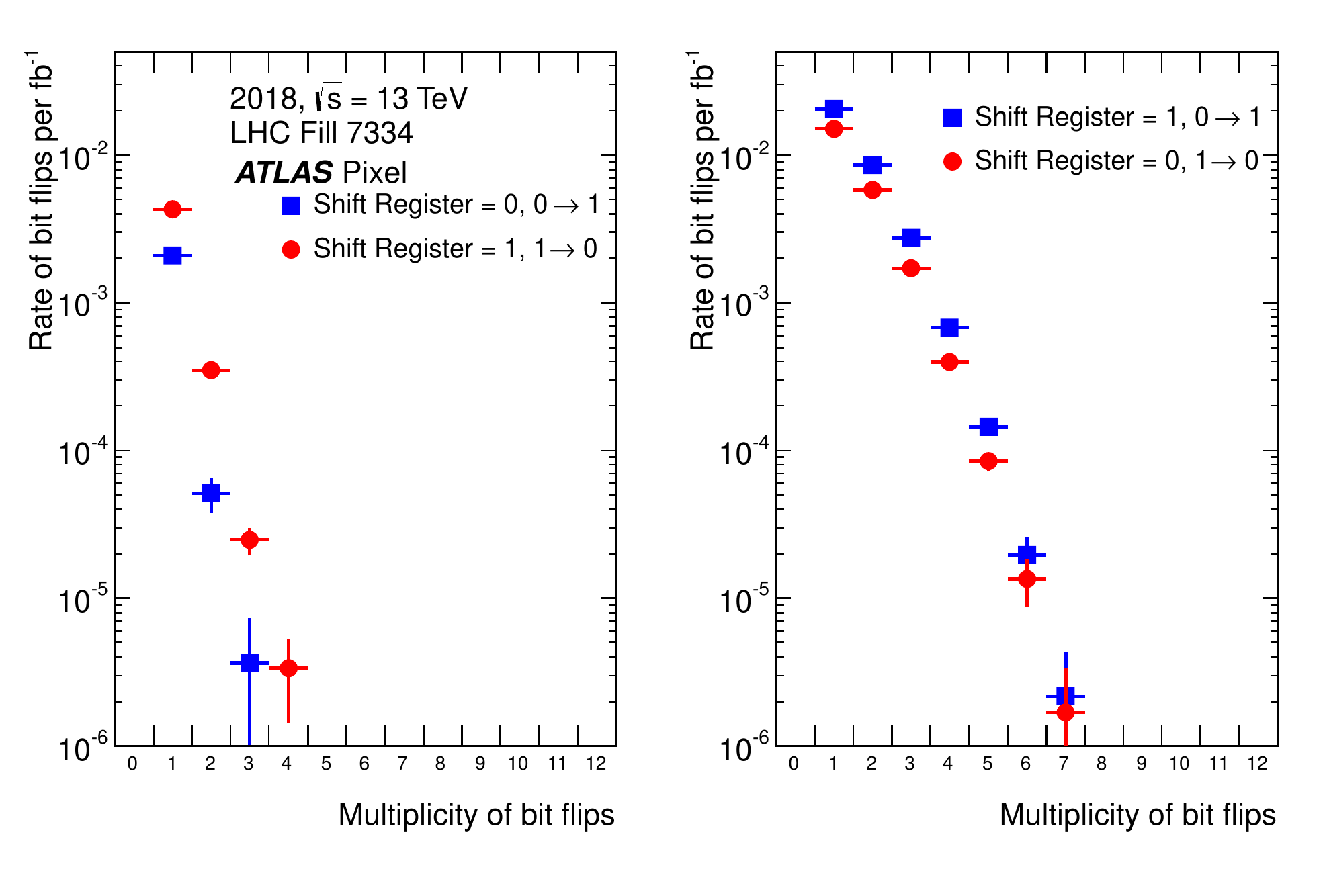}
\put(25,-1){(a)}
\put(75,-1){(b)}
\end{overpic}
\caption{
Distribution of SEU/SET bit flip multiplicity  in pixel memory of FE-I4B. In figure (a) the Shift Register is set to the same value as the memory, while in figure (b) the state of the Shift Register was set to the opposite of the memory content. 
}
\label{fig:hn4mult}
\end{figure}
Figure~\ref{fig:hn4mult} shows the distribution of SEU/SET bit flip multiplicity in pixel memory of FE-I4B
for the fill 7334 (integrated luminosity \SI{0.35}{\per\femto\barn},
peak luminosity $1.1 \times 10^{34}$ cm$^{-2}$ s$^{-1}$).
In figure~\ref{fig:hn4mult} (b), the Shift Register state was set to the opposite of the memory
 content, and bit flips are mainly due to SET on the LOAD line. 
 Each pixel has 13 local memory bits, so multiple bit flips per pixel could be a sign of glitches
 on the common LOAD line.
For example, the glitch generated on the LOAD line before the fan-out to 
13 LOAD lines of DICE latches (see figure~\ref{fig:dice_scematics})
may result in the synchronous bit flip of many bits (up to 13). 
 The distributions are peaked at one bit flip per pixel,
 which indicates that the glitch on the LOAD line is mainly specific to each DICE latch bit.
 However the long tail in the multiplicity distribution may be a sign of small contributions from glitches common to several bits.
In figure~\ref{fig:hn4mult} (a), the Shift Register is set to the same value as the memory,
 so the bit flips are mainly due to true SEU.
  
\subsection{Global register configuration read-back measurements ~\label{sec:globalreadback}}
\begin{figure}[h]
\centering
 \begin{overpic}[width=0.82\linewidth]{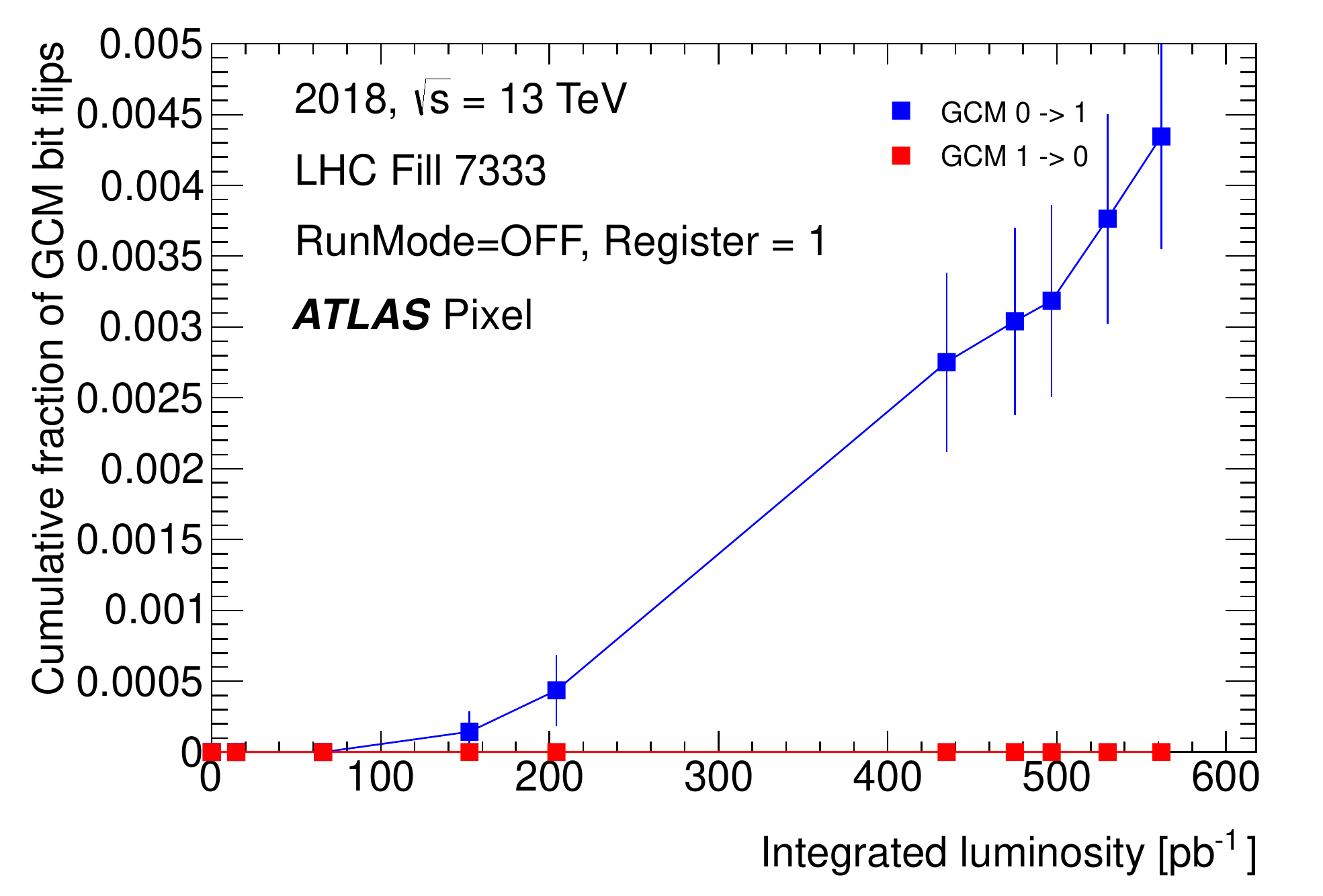}
\end{overpic}
\caption{
Cumulative fraction of SEU/SET bit flips in the Global Configuration Memory
 (GCM) of FE-I4B as a function of integrated luminosity in LHC fill 7333.
}
\label{fig:grseuvslumi}
\end{figure}
During LHC fill 7333 the values of the Global Configuration Memory were 
read back several times and compared with the initial setting.
In figure~\ref{fig:grseuvslumi}, the cumulative fraction of SEU/SET bit flips in the Global Configuration Memory of FE-I4B is shown as a function of integrated luminosity.
 The high rate of $0\rightarrow1$ flips is probably due to SET on the LOAD line
 with register value 1.
The pattern 0xFFFF was loaded last into the register.
The values of the register are refreshed several
 times during the fill. No $1\rightarrow0$ real SEU transitions were observed during
 the fill due to the triply redundant memory design.

\subsection{SEE cross sections ~\label{sec:crosssections}}
For the calculation of SEE cross sections, we use the average prediction of
the flux of hadrons with energy above 20 MeV
\footnote[5]{The choice of 20 MeV hadron energy cut was justified in
~\cite{ref:Huhtinen} by effective reduction of SEU effects below this energy.}
in the two outer 3D IBL modules: $91.0\times 10^{11}$cm$^{-2}$fb$^{-1}$ ~\cite{ref:IanDawson}.
This flux was calculated from 13 TeV proton-proton minimum bias events generated using PYTHIA8~\cite{ref:PYTHIA}, applying the
A3
\footnote[6]{A3 is one of the tunes of PYTHIA8 Monte-Carlo generator
to the ATLAS LHC data.}
 tune to ATLAS data, and using NNPDF23LO Parton Density Functions (PDF) ~\cite{ref:Dawson}. The flux of hadrons in the
ATLAS detector was simulated using FLUKA version 2011.2x particle transport
code~\cite{ref:FLUKA}. The systematic error on the hadron flux was estimated
to be $\sim30\%$ ~\cite{ref:IanDawson}.

\begin{table}[htbp]
\centering
\begin{tabular}{|l|l|l|l|l|}
\hline
Beam, SEE type & Transition & SR       & Rate per fb$^{-1}$ [\%] & Cross-section\\
               &            & value    & (stat./syst.)     & 10$^{-15}$ cm$^2$\\
\hline
LHC: mainly SEU&  $0\rightarrow1$ & 0  & 0.22 $\pm$ 0.01 $\pm$ 0.09     & 0.24 $\pm$ 0.13\\
\hline
LHC: mainly SEU&  $1\rightarrow0$ & 1  & 0.46 $\pm$ 0.01 $\pm$ 0.19     & 0.51 $\pm$ 0.26\\
\hline
LHC: mainly SET&  $1\rightarrow0$ & 0  & 3.07 $\pm$ 0.02 $\pm$ 0.80     & 3.39 $\pm$ 1.34\\
on LOAD line   &                  &               &                 &      \\
\hline
LHC: mainly SET&  $0\rightarrow1$ & 1  & 4.68 $\pm$ 0.03 $\pm$ 1.21     & 5.16 $\pm$ 2.04\\
on LOAD line   &                  &               &                 &      \\
\hline
24 GeV protons &  $0\rightarrow1$ & 0  & n.a.                           & 1.10 \\
Mostly SEU with &                  &               &                 &      \\
some SET admixture &                  &            &                    &      \\
\hline
\end{tabular}  
\caption{Summary of measured rates and SEE cross sections.}
\label{tab:abscros}
\end{table}

In table~\ref{tab:abscros}, we summarize the measured rates and SEE cross sections.
The rates presented are averages over all available chips and 12 pixel
memory bits, excluding the HitBus bit (which shows abnormal rates
related to the special layout interconnecting all pixels in the matrix).
The systematic errors on the SEE rates were calculated from the RMS spread
of the rates in different IBL chips and in different pixel memory bits for SEU
and SET dominated samples. SET cross sections are greater than SEU 
cross sections by a factor of 10.
Some differences in $0\rightarrow1$ and $1\rightarrow0$ transitions are observed, but fall within the present
systematic errors. These variations are related to differences in the geometrical positions of the chips,
differences in process variations, and differences in the mixture	of SEU and
SET effects. 

The measured cross sections are compatible with the cross sections
measured with 24 GeV protons at CERN PS. So for future HL LHC experiments,
measurement of SEE cross sections using the 24 GeV proton PS beam is recommended, and special attention should be paid to the
optimization of chip design not only against SEU, but also against SET effects.

%% file: documents/recovery.tex
\label{sec:recovery}
Data that is being assembled into an event fragment from multiple sources becomes de-synchronized when the constituent components actually belong to different events.
Such de-synchronization can happen at module level (between FEs), or at readout level (between modules). For this reason, every \SI{5}{\second}, the ATLAS Central Trigger Processor (CTP) sends
an Event Counter Reset signal (ECR) inside a \SI{2}{\milli\second} time window without triggers. The ECR command gets propagated from the various sub-detector readout systems to the corresponding FEs in order
to regularly re-synchronize the entire readout chain.
The IBL DAQ system exploited the \SI{2}{\milli\second} window to perform regular reconfigurations of the FE-I4B registers; only the global registers were reloaded regularly, while the reconfiguration of single pixel registers was only tested in a few runs. A short description of the software/firmware procedure is described in sections ~\ref{subsec:GRRecon} and ~\ref{subsec:PRRecon}.

\subsection{Global register reconfiguration}
\label{subsec:GRRecon}
\begin{figure}[h]
\centering
 \begin{overpic}[width=1.00\linewidth]{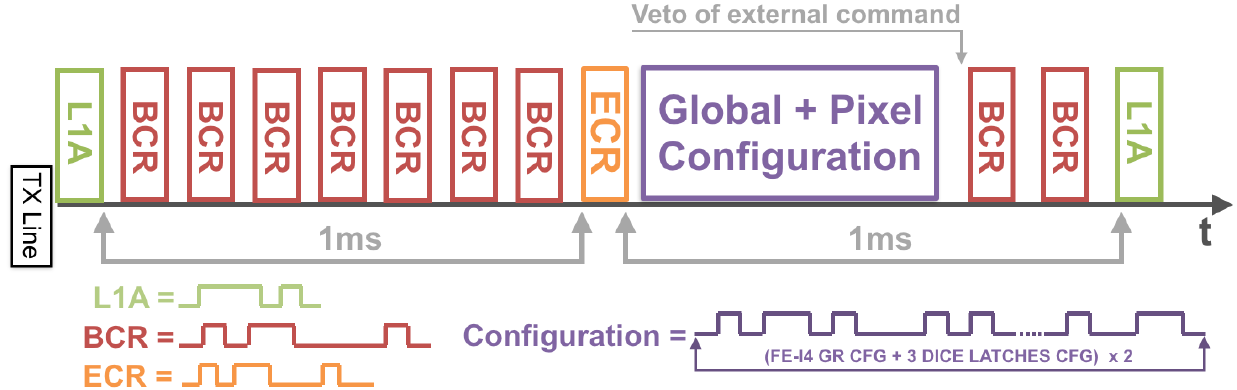}
\end{overpic}
\caption{
Reconfiguration mechanism at ECR timing for refreshing of global and pixel registers.
}
\label{fig:IBLGRReconfig}
\end{figure}
 The first implementation of periodic reconfiguration every 5 seconds on the IBL FEs was officially released in ATLAS data taking in August 2017,
when the instantaneous luminosity was up to $1.6 \times 10^{34}$\,cm$^{-2}$\,s$^{-1}$.
In this version, only the global registers of the IBL FEs are reloaded. A full schematic of the reconfiguration action is shown in figure~\ref{fig:IBLGRReconfig}.

This action is taken as soon as an ECR signal sent by the CTP reaches the IBL readout electronics. In particular, each IBL ROD is responsible for detecting the ECR signal and propagating it downstream to the corresponding portion of IBL FEs.
During the next 1 \SI{}{\milli\second} period the ROD inhibits the actions of the CTP signals on the FE-I4B serial lines,
including the Bunch Counter Reset signal (BCR), which is broadcast by the CTP every LHC orbit (11 kHz frequency).
At the same time, the BOC card, which provides the optical communication interface between the ROD and the modules, detects the ECR and appends to it a sequence of configuration commands, one for each of the 36 FE-I4B global registers. The BOC card has the ability to store the values of each single GR for each FE-I4B in its FIFOs, and it can send a customized stream to all of the modules in parallel. The on-ROD software (running in the PPC), must load the BOC FIFOs at the beginning of each run or whenever there is a change in the module configuration.

The entire procedure of reconfiguration must not last more than \SI{1}{\milli\second}, since the ATLAS CTP will start triggering again after this time interval.
A total of 110 \SI{}{\micro\second} is needed to complete the global register reconfiguration.

The main positive effects of the the global register reconfiguration (LV current and hit occupancy stability) have been described in section~\ref{sec:intro} and are visible in figure~\ref{fig:SEUEffectsGlobal}. Furthermore, this procedure significantly reduces the dead time in ATLAS caused by FE-I4B misconfiguration.
\begin{figure}[h]
\centering
 \begin{overpic}[width=0.49\linewidth]{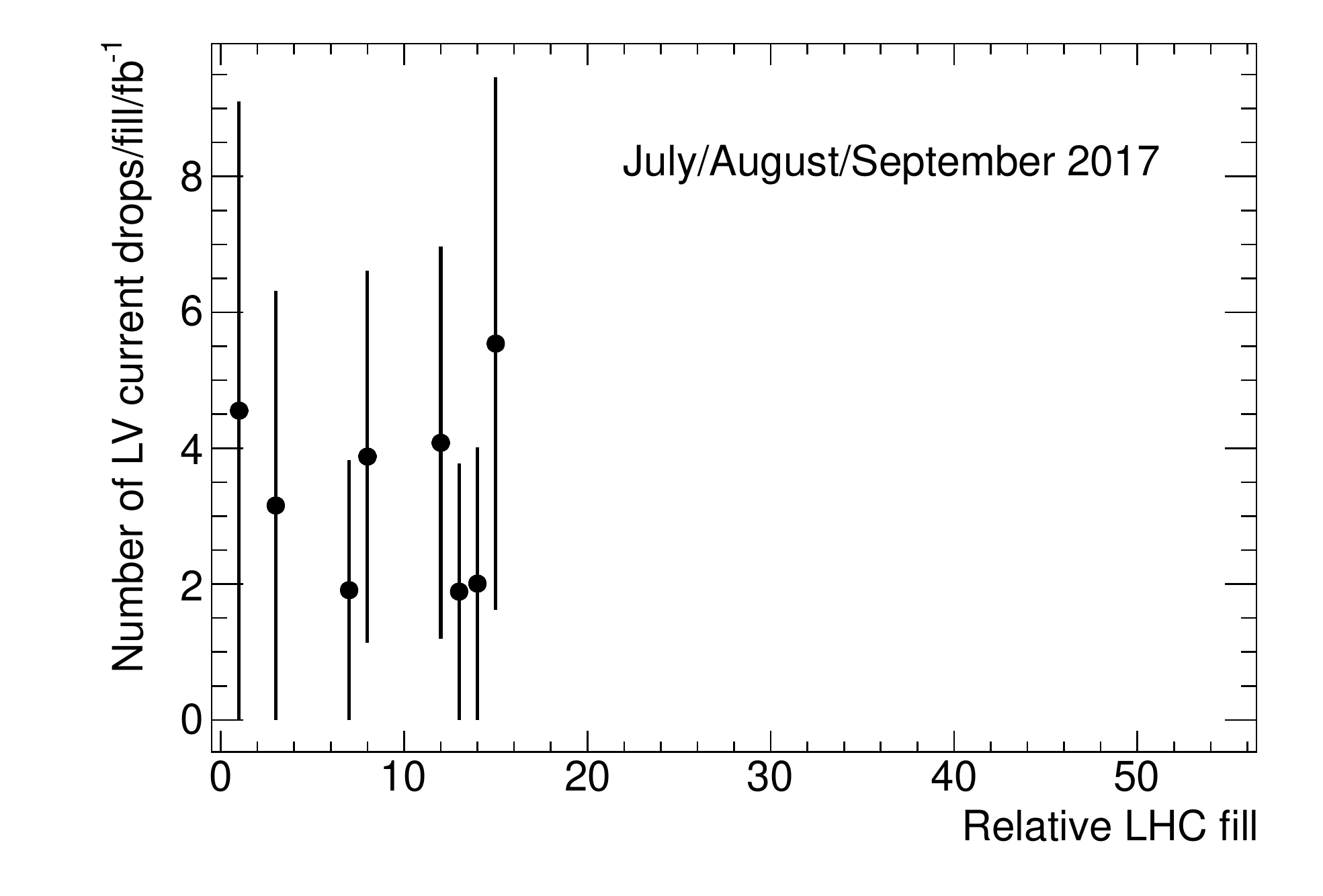}
\put(50,-2){(a)}
\end{overpic}
 \begin{overpic}[width=0.49\linewidth]{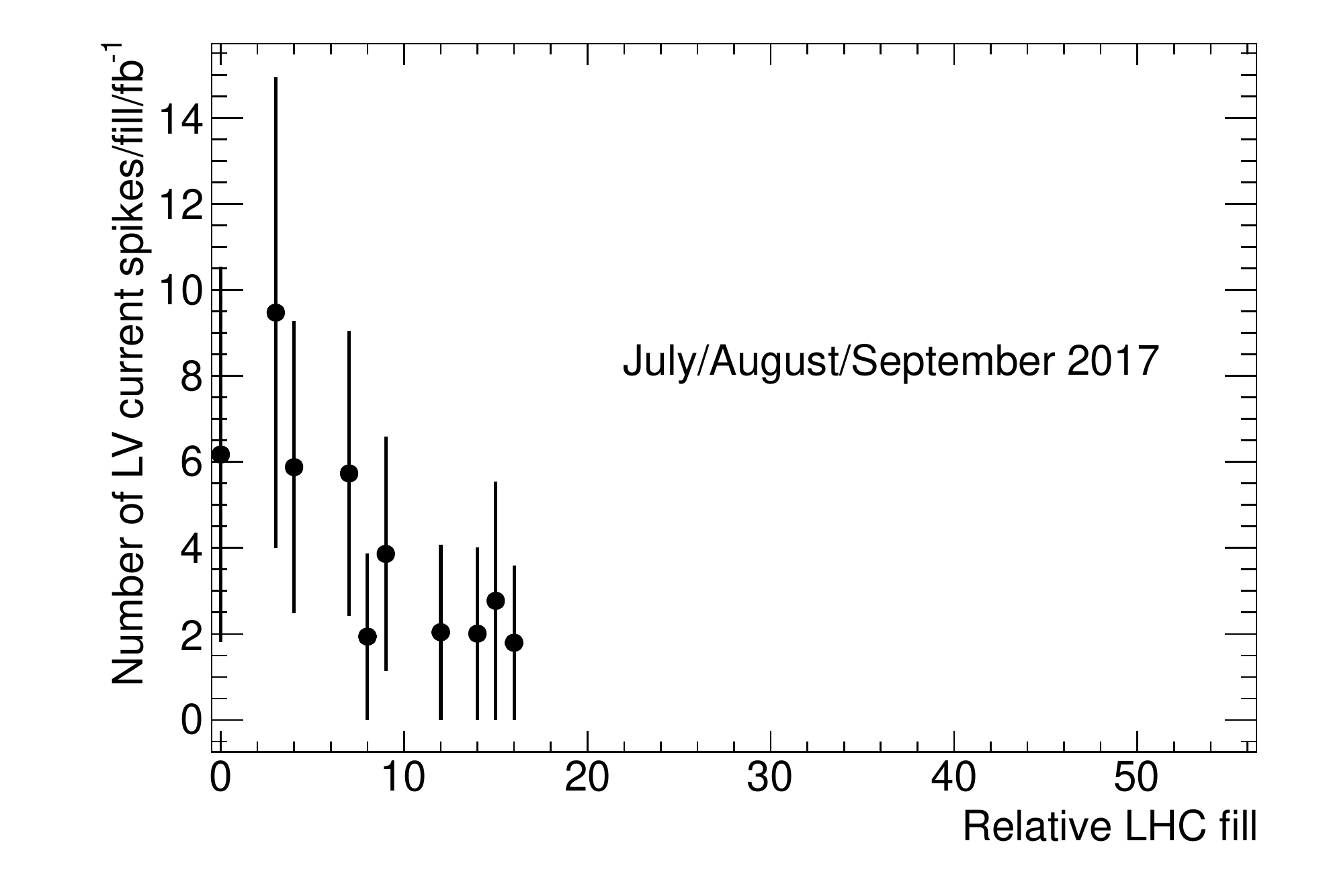}
\put(50,-2){(b)}
\end{overpic}
\vspace{0.15cm}
\caption{
Number of low voltage current drops (a) and spikes (b) per fb$^{-1}$ in LHC
fills in July through September of 2017. 
Relative LHC fill number is shown on the horizontal axis.
Starting from relative fill number 22 (LHC fill 6090), the reconfiguration of
the FE-I4B was implemented in all staves, and no drops (spikes) are observed from this fill forward. 
In relative fills 11-21 (LHC fills from 6044 to 6089), a test of the FE-I4B reconfigurations was performed
in some parts of the IBL. This resulted in more fills without current drops (spikes) compared to the earlier 
fills when no regular reconfiguration was implemented. 
}
\label{fig:ILVDrops}
\end{figure}

Following the implementation of the global register reconfiguration procedure, the occurrence of drops and spikes in the LV current is dramatically reduced.  Figure~\ref{fig:ILVDrops} shows that no drops or spikes are observed after relative fill number 22 (LHC fill 6090, 16 August 2017), when the procedure was fully implemented.
\interfootnotelinepenalty=1000
\footnote[7]{Current drops (or spikes) of 1.5\% in four chip modules are counted if they last longer than 100 seconds and occur during LHC fills longer than four hours. The first and the last half hour in each fill are excluded to avoid edge instabilities. To suppress current drops (or spikes) due to ROD configurations and luminosity fluctuations, we also require the drop occur in only one IBL module during in a $\pm$ 20 min interval.}

\subsection{Single pixel latch reconfiguration} 
\label{subsec:PRRecon}
The continuous FE-I4B global memory reconfiguration at ECR has been demonstrated to be very effective for ATLAS data taking. For this reason, the regular reconfiguration of single pixel latches was also investigated. 
Unfortunately, there is not enough time in the ECR time window to reconfigure all pixel latches at once.

The solution implemented addresses up to three latches of one pixel double column at each ECR.
In total the FE-I4B hosts 13 pixel latches and 40 double columns; only 11 out of 13 latches (Enable, HitBus, TDAC and FDAC) are regularly reconfigured because of their direct impact on noise and quiet pixels.
Any relevant pixel latch is reconfigured every $\sim$130 ECRs (meaning a full reconfiguration every $\sim$11 minutes).
The HitBus latch is actually refreshed more often than others, since a uniform value must be loaded in the entire chip;
this bit can indeed be loaded in parallel in all the pixels using a different register writing modality.
Unlike the global reconfiguration, a different BOC FIFO content is required at each ECR, since a different group of latches and a different double column are addressed every time.
This implementation required a more intense use of the on-ROD PPC. The automatic reconfiguration of the pixel latches was carried out only in few test runs in July 2018 for a relevant portion of the detector (all the IBL FEs located at $\eta=\pm\,2.8$, corresponding to IBL DAQ modules in position A7 and C7).
Unfortunately, due to some instabilities in the on-ROD PPC software, this implementation was only used for test runs and was not fully deployed for the remaining LHC Run 2 period.

\begin{figure}[h]
\centering
 \begin{overpic}[width=0.49\linewidth]{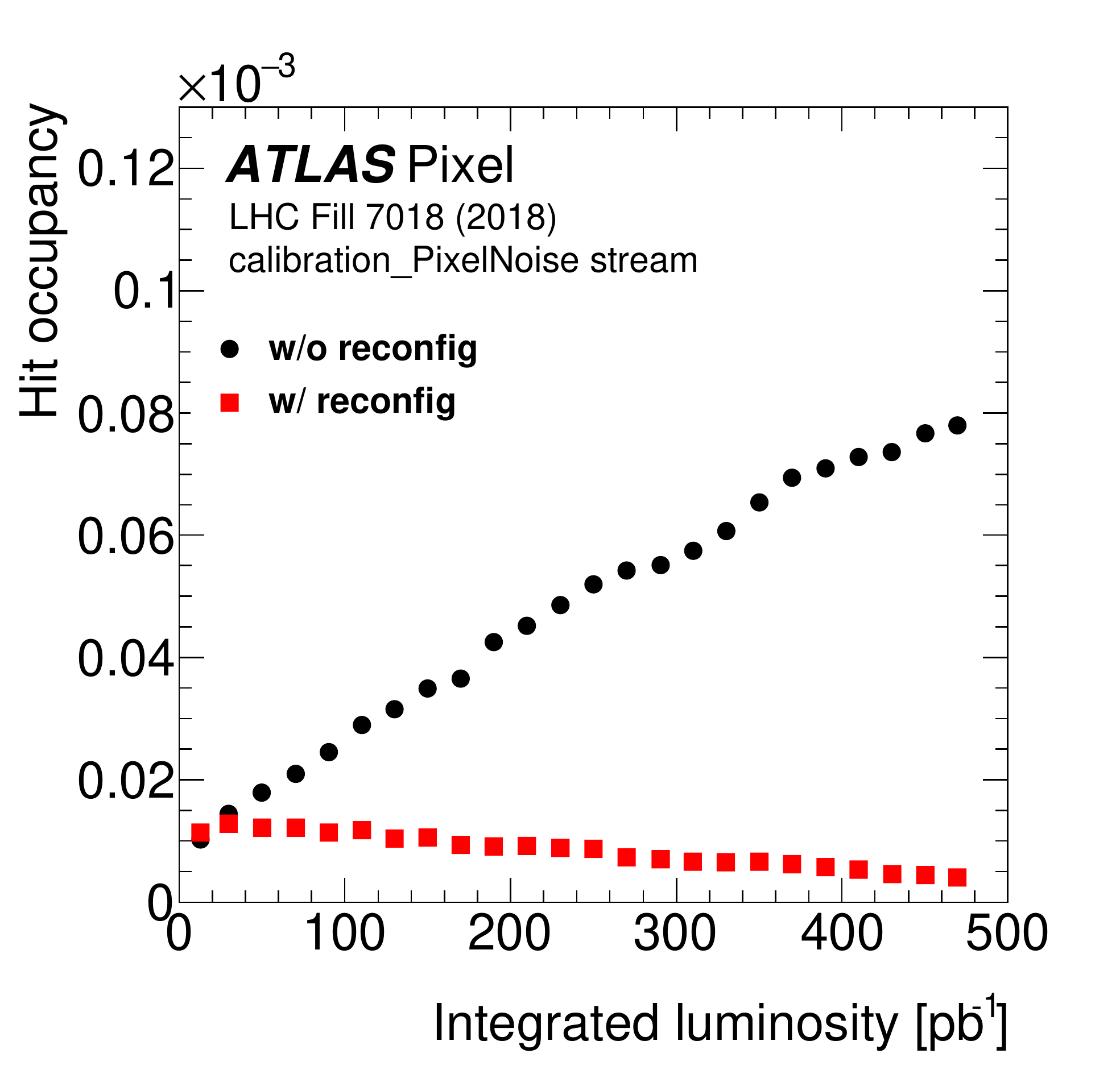}
\end{overpic}
\caption{
The noise occupancy in IBL
3D modules as a function of integrated luminosity with
and without the mechanism of pixel register
reconfiguration at ECR.
}
\label{fig:yosuke}
\end{figure}
Figure~\ref{fig:yosuke} shows the noise occupancy (the
number of hits per pixel per event) in IBL
3D modules as a function of integrated luminosity with
and without the mechanism of pixel register
reconfiguration at ECR timing. The
mechanism is adopted for modules at A7
and C7 and is not deployed for those at A8
and C8.
Figure~\ref{fig:yosuke} is made from the
calibration PixelNoise stream, where random data
are taken with empty, isolated bunches.
The data points include statistical error bars. Without the
mechanism of pixel register reconfiguration,
SEUs cause an increase in the noise occupancy during the
run. When the reconfiguration mechanism is adopted,
the noise decreases as a function of
integrated luminosity. This is
because the reconfiguration is performed
every 11 minutes for the same pixel registers,
but the time required to accumulate the same amount of
integrated luminosity becomes longer during
the run due to the decrease of  instantaneous
luminosity.

\begin{figure}[h]
\centering
 \begin{overpic}[width=0.49\linewidth]{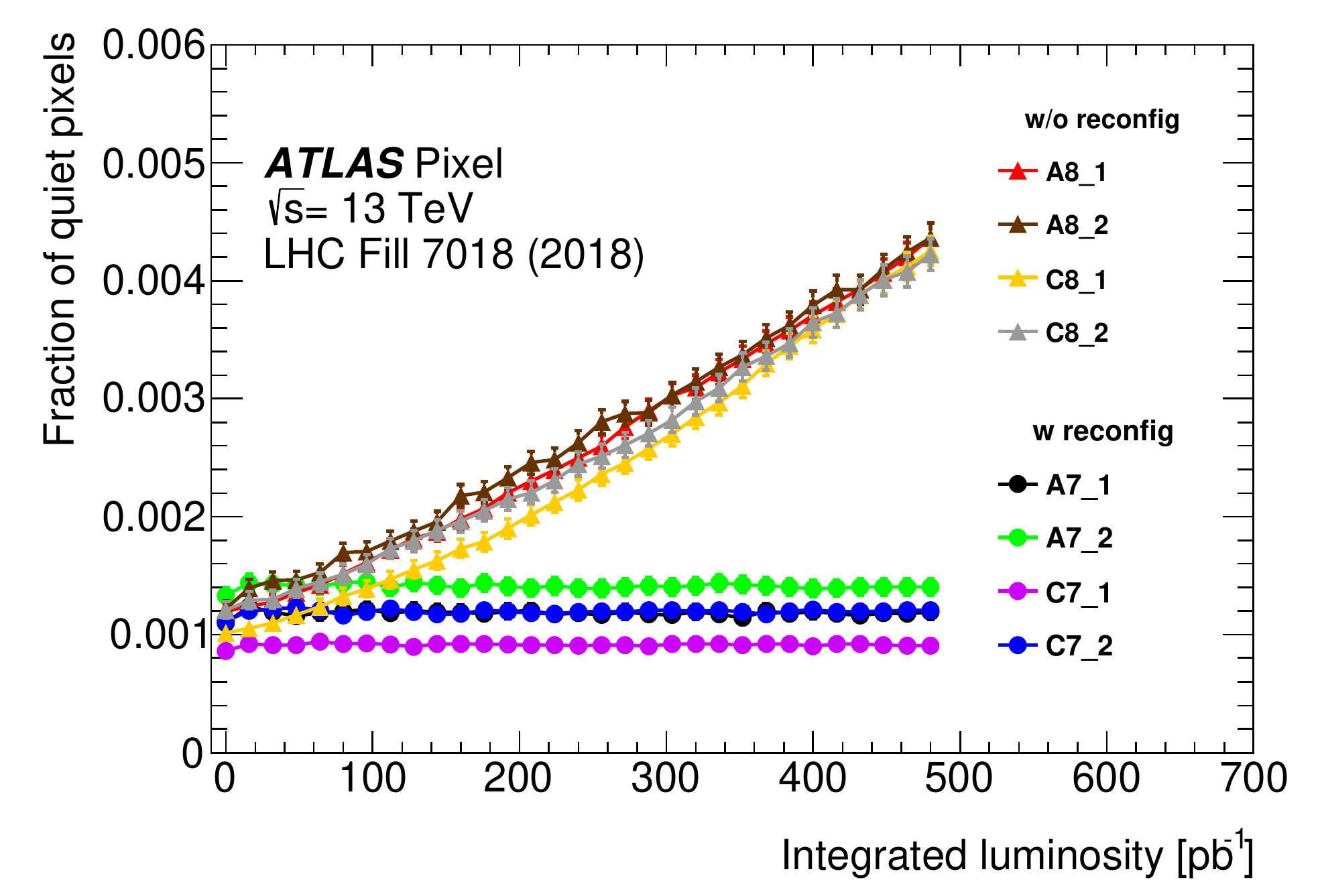}
\put(50,-5){(a)}
\end{overpic}
 \begin{overpic}[width=0.49\linewidth]{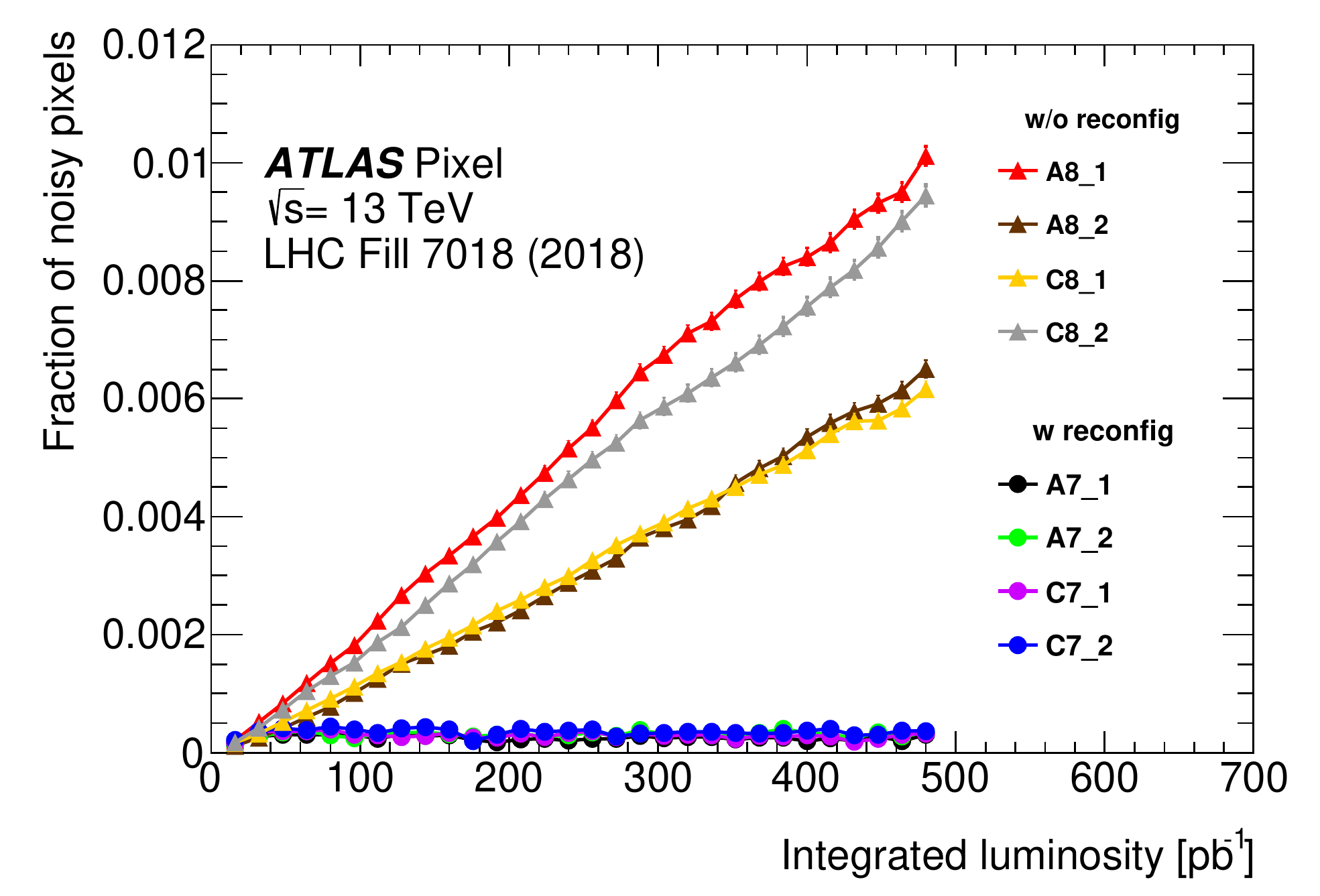}
\put(50,-5){(b)}
\end{overpic}
\vspace{0.15cm}
\caption{
 Fraction of quiet (a) or noisy (b) pixels versus integrated luminosity in fill 7018 from 2018, shown in the eight 3D IBL $\eta$ rings. Long pixels and pixels in the region of L-shape noise were not used in the analysis.}
\label{fig:noisy-quiet-fill7018}
\end{figure}

 Figure~\ref{fig:noisy-quiet-fill7018} (a) ((b)) shows the fraction of quiet (noisy) pixels
as a function of integrated luminosity
 in the eight IBL $\eta$ rings in the LHC fill 7018 of 2018.
 The data are extracted from randomly triggered events, as discussed
in section \ref{subsec:clusterdata}.  
 In the modules without regular reconfiguration, more pixels become noisy or quiet during data taking due to SEU. In the modules reconfigured every 11 minutes, the fraction of noisy or quiet pixels is more stable.

\begin{figure}[h]
\centering
 \begin{overpic}[width=0.74\linewidth]{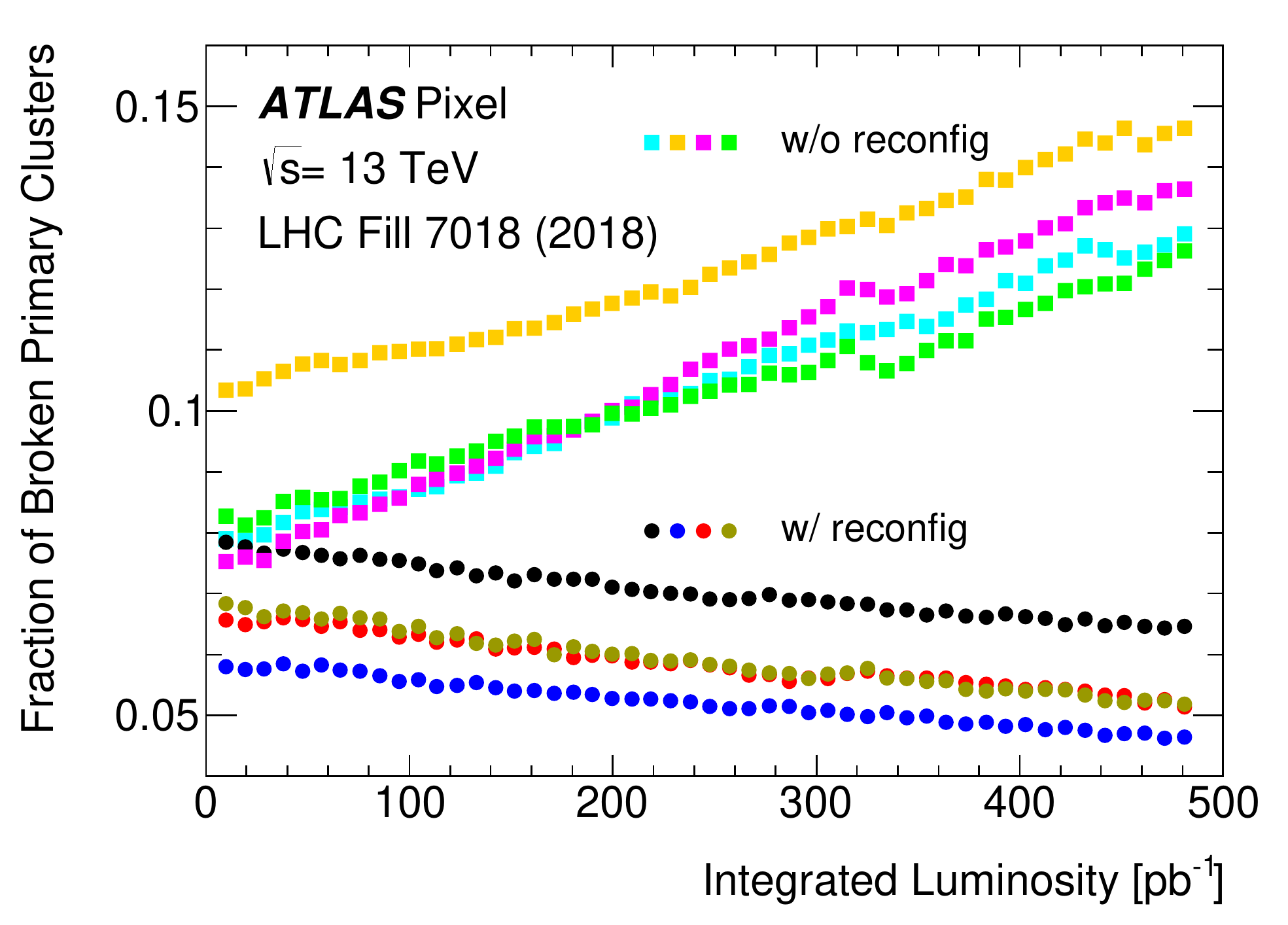}
\end{overpic}
\caption{Fraction of broken primary clusters versus integrated luminosity in fill 7018 from 2018, shown in the eight rings of 3D modules.}
\label{fig:broken-fill7018}
\end{figure}
Figure~\ref{fig:broken-fill7018} shows the fraction of broken primary clusters
as a function of integrated luminosity
 in fill 7018 from 2018, in the eight rings of 3D modules. The 14 modules in the same ring have been studied together to achieve higher statistics. The four rings of 3D modules without regular reconfiguration show a similar increase in the rate of broken clusters. The four rings of 3D modules with regular reconfiguration show a similar decrease in the rate of broken clusters. This is because the full reconfiguration occurs every 11 minutes, and more time is needed to accumulate the same amount of luminosity at later times when the instantaneous luminosity is lower. The modules on IBL stave 6 were reconfigured after $\sim$320\,pb$^{-1}$ of data were collected. This corresponds to the changes of slope in the modules without reconfiguration.

\begin{figure}[h]
\centering
 \begin{overpic}[width=0.49\linewidth]{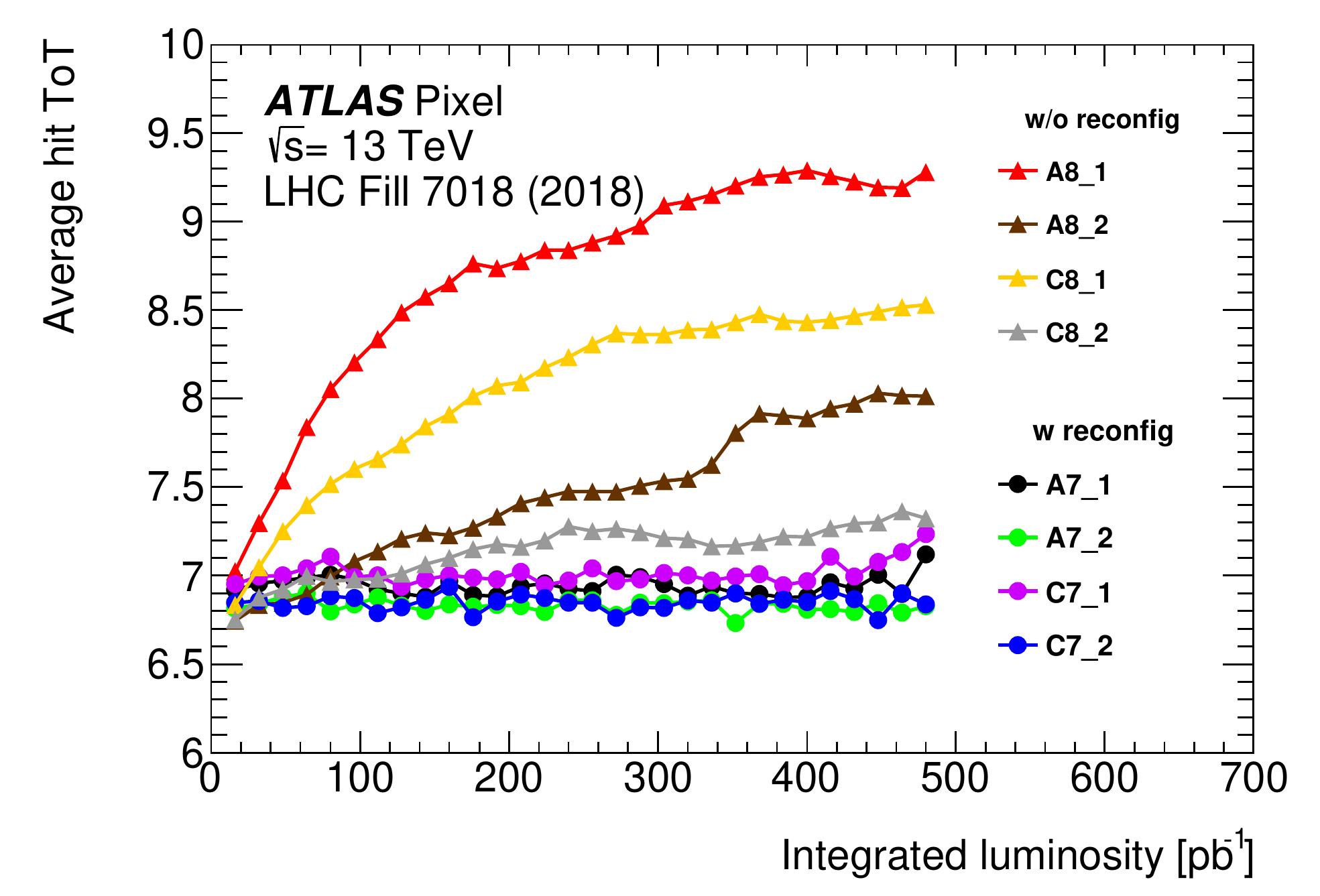}
\put(50,-5){(a)}
\end{overpic}
 \begin{overpic}[width=0.49\linewidth]{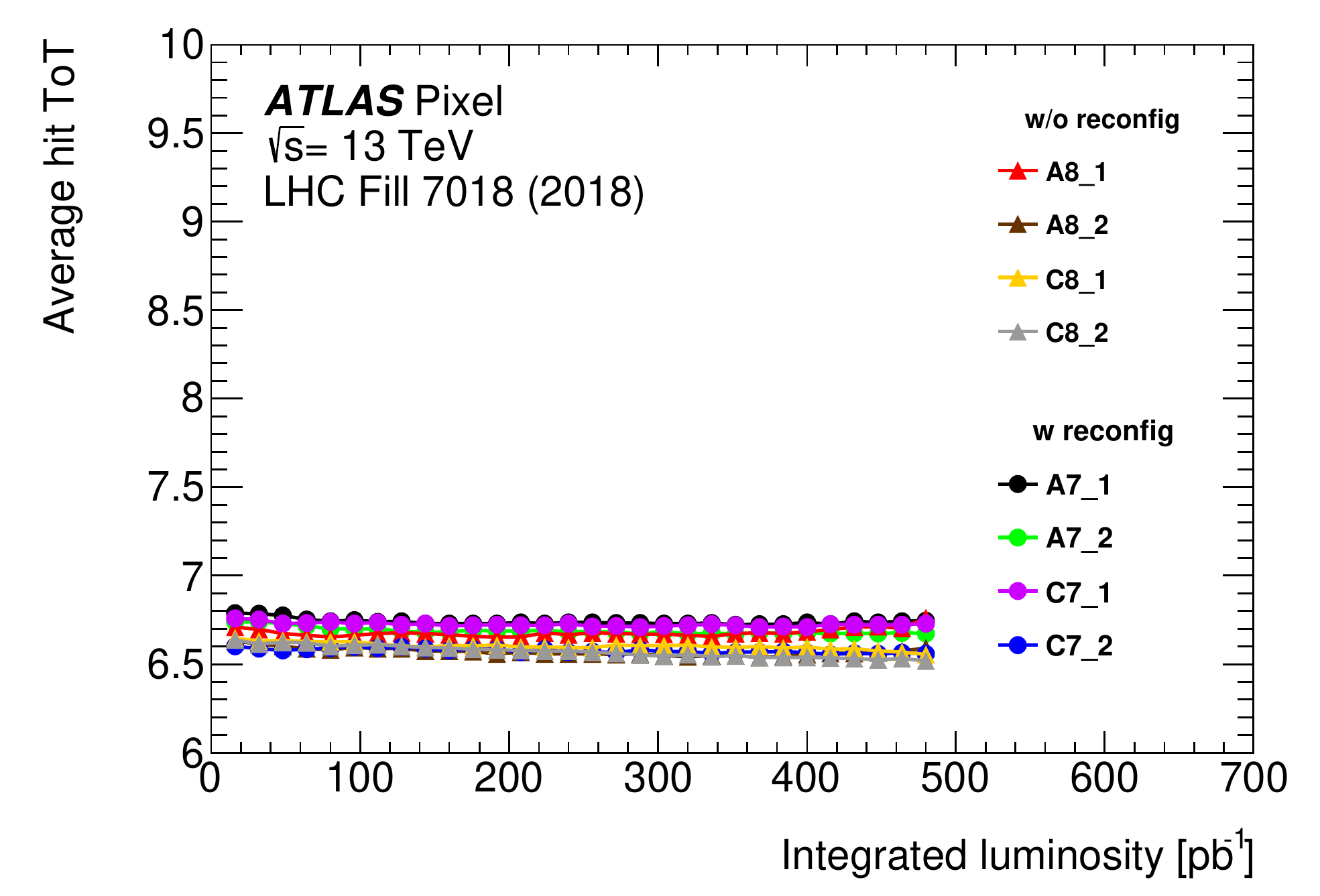}
\put(50,-5){(b)}
\vspace{0.15cm}
\end{overpic}
\vspace{0.15cm}
\caption{
(a) Average hit ToT of fired pixels from single-pixel clusters for
 eight 3D IBL $\eta$ rings versus integrated luminosity.
(b) Average hit ToT of fired pixels from single-pixel clusters
for eight $\eta$ rings, excluding noisy pixels.}
\label{fig:ToT-fill7018}
\end{figure}

Figure~\ref{fig:ToT-fill7018} (a) shows the average hit ToT
 from single-pixel clusters as a function of integrated luminosity
 in the LHC fill 7018 from 2018, in the eight 3D IBL $\eta$ rings.
In the modules without regular reconfiguration, the average ToT increases 
because more pixels become noisy due to SEU. In the modules reconfigured every
11 minutes, the average ToT is more stable. To confirm that the increase of ToT is due to noisy pixels, the average hit ToT is studied with noisy pixels excluded (noisy pixels are identified in figure~\ref{fig:noisy-quiet-fill7018} (b)).
 Figure~\ref{fig:ToT-fill7018} (b) shows that the average ToT in the modules without regular reconfiguration is more stable after excluding noisy pixels, which proves that the increase of ToT during the fill is almost fully correlated with noisy pixels caused by SEU in TDAC MSB. The detuning of pixels due to SEU in FDAC causes the ToT to get smaller or larger.
Due to the dominance of 0$\rightarrow$1 transitions we expect an increase in FDAC values more often than a decrease, but the effect is small and not
visible on figure~\ref{fig:noisy-quiet-fill7018} (b).

\begin{figure}[h]
\centering
 \begin{overpic}[width=0.72\linewidth]{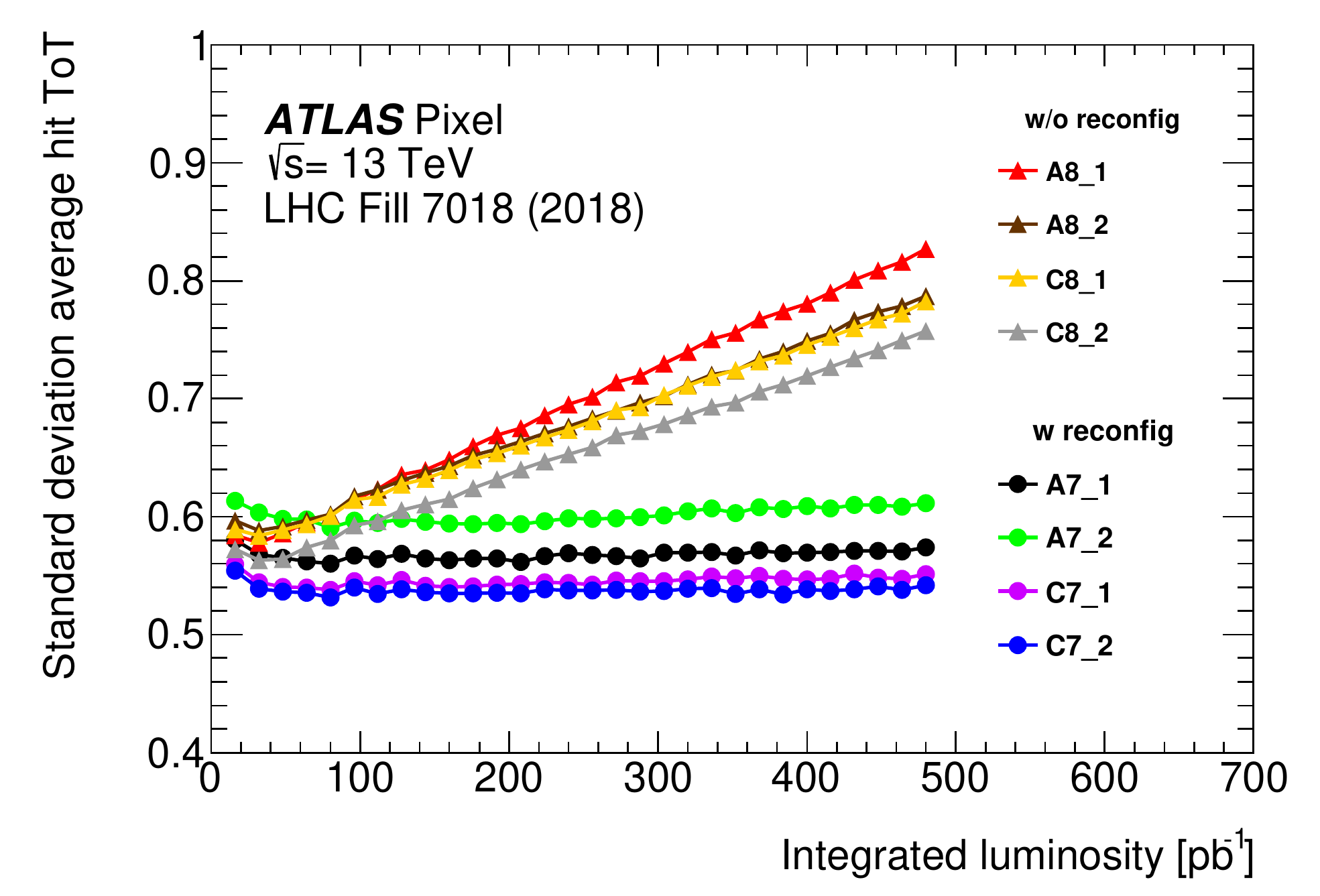}
\end{overpic}
\caption{Standard deviation of the average ToT per pixel after noise subtraction
versus integrated luminosity in fill 7018 from 2018, shown in the eight rings of 3D modules.}
\label{fig:totrms-fill7018}
\end{figure}
Figure~\ref{fig:totrms-fill7018} shows the standard deviation of the average ToT
per pixel after noise subtraction versus integrated luminosity in
 the LHC fill 7018 from 2018, shown in the eight rings of 3D modules.
In the four rings without reconfiguration we observe the increase
of the standard deviation, probably due to the flips of FDAC bits.

%% file: documents/conclusion.tex
   The ATLAS Pixel detector, including the IBL,
was operated efficiently
at high luminosity, and SEUs were observed at rates close to the expectation based on previous tests in 24 GeV proton beam.
SEUs in the global and local pixel configuration memories of FE-I4B chip
were observed during LHC fills and had several consequences: module de-synchronization,
current jumps, dead modules, quiet pixels, noisy pixels, broken clusters, etc.
By read-back measurements and simulations of the electronics, it was demonstrated that SETs on the LOAD line of the 
DICE latch dominate the memory flips.
These SETs explain the apparent asymmetries of $0\rightarrow1$ and $1\rightarrow0$ transitions.

Starting in 2016, some of the effects of SEU/SET were mitigated
by occasional manual
or automatic reconfiguration of problematic modules.
The impact of SEUs on the global configuration was mitigated in 2017
by refreshing the global memory
every 5 seconds; this was syncronized with ECR and therefore
resulted in no additional dead time to ATLAS.
It was demonstrated in a few test runs in 2018 that gradually refreshing local pixel memory during ECR reduces the number of noisy and quiet pixels and broken clusters.
We plan to introduce this local memory refreshing in all IBL modules in 2021 for LHC Run 3
and therefore improve the performance with potential future increases of the LHC luminosity.

The operational SEU/SET experience with FE-I4B was also used to inform the
 design of the next generation of LHC pixel front end chip FE-RD53
 ~\cite{ref:RD53requirements}, ~\cite{ref:RD53manual}.
 High frequency refreshing of all global and local memories (sometimes referred to as trickle configuration) is embedded in the design of this chip. The RD53 command protocol permits sending configuration and 
sub-circuit reset instructions concurrently with data acquisition triggers,
without incurring dead time.

%% file: acknowledgements/ANA-PIXE-2019-02-PAPERAcknowl.tex


We acknowledge the support of  
NSERC, NRC and CFI, Canada; 
CERN; 
MSMT CR, MPO CR and VSC CR, Czech Republic; 
IN2P3-CNRS, CEA-DSM/IRFU, France; 
BMBF, HGF, and MPG, Germany; 
INFN, Italy; 
MEXT and JSPS, Japan; 
NWO, Netherlands; 
RCN, Norway; 
MES of Russia and NRC KI, Russian Federation; 
ARRS and MIZ\v{S}, Slovenia; 
MINECO, Spain; 
SRC and Wallenberg Foundation, Sweden; 
SERI, SNSF and Cantons of Bern and Geneva, Switzerland; 
MOST, Taiwan; 
STFC, United Kingdom; 
NSF MRI award PHY-1039175 and DOE, United States of America. 


We are very grateful to Dave Robinson for his contribution as a member of Editorial Board of this paper. We are grateful to Alessandro Polini for careful reading of the paper.

We acknowledge CERN and the ATLAS Collaboration, and in particular the ATLAS TC (Technical Coordination) and CERN EP-DT groups, for their support during the maintenance and operation of the ATLAS Pixel detector. We further acknowledge the support staff from participating Pixel institutes, who were essential to the success of the project.